\documentclass[11pt]{article}
\usepackage[utf8]{inputenc}

\usepackage[bookmarksnumbered,colorlinks,citecolor=red,linkcolor=red,hyperindex,linktocpage=true]{hyperref}
\usepackage{authblk} 

\usepackage{amsthm}
\usepackage{amssymb}
\usepackage{amsmath}
\usepackage{bm,xcolor}
\usepackage{physics}
\usepackage{geometry}
\usepackage{fullpage}

\usepackage{graphicx,subfigure,epstopdf,float}

\renewcommand{\i}{\ensuremath{\text{\normalfont I}}}
\newcommand{\ii}{\ensuremath{\text{\normalfont I\!I}}}

\newcommand{\ici}{\ensuremath{\text{\normalfont I,I}}}
\newcommand{\icii}{\ensuremath{\text{\normalfont I,I\!I}}}
\newcommand{\iici}{\ensuremath{\text{\normalfont I\!I,I}}}
\newcommand{\iicii}{\ensuremath{\text{\normalfont I\!I,I\!I}}}

\newcommand{\iiciii}{\ensuremath{\text{\normalfont I\!I,}\Gamma}}
\newcommand{\iiicii}{\ensuremath{\Gamma,\text{\normalfont I\!I}}}

\newtheorem{theorem}{Theorem}[section]

{\bf}{\it}

\title{Quantum Simulation  for  Quantum Dynamics with Artificial Boundary Conditions}
\title{Quantum Simulation  for  Quantum Dynamics with Artificial Boundary Conditions}
\author[1,2]{Shi Jin\thanks{shijin-m@sjtu.edu.cn}}
\author[1, 2, 3]{Nana Liu\thanks{nana.liu@quantumlah.org}}
\author[4]{Xiantao Li\thanks{xxl12@psu.edu}}
\author[1]{Yue Yu\thanks{terenceyuyue@sjtu.edu.cn}}
\affil[1]{School of Mathematical Sciences, Institute of Natural Sciences, MOE-LSC, Shanghai Jiao Tong University, Shanghai, 200240, P. R. China.}
\affil[2]{Shanghai Artificial Intelligence Laboratory, Shanghai, China.}
\affil[3]{University of Michigan-Shanghai Jiao Tong University Joint Institute, Shanghai 200240, China.}
\affil[4]{Department of Mathematics, The Pennsylvania State University, University Park, Pennsylvania 16802, USA}

\begin{document}

\maketitle

\begin{abstract}
    Quantum dynamics, typically expressed in the form of a time-dependent Schr\"odinger equation with a Hermitian Hamiltonian,  is a natural application for quantum computing.  However,
    when simulating  quantum dynamics that involves the emission of electrons, it is necessary to use artificial boundary conditions (ABC) to confine the computation within a fixed domain.  The introduction of ABCs alters the Hamiltonian structure of the dynamics, and existing quantum algorithms can not be directly applied since the evolution is no longer unitary. The current paper utilizes a recently introduced Schr\"odingerisation method \cite{jin2022quantumshort, jin2022quantum}  that converts non-Hermitian dynamics to a Schrödinger form, for the artificial boundary problems.   We implement this method for three types of ABCs, including the complex absorbing potential technique, perfectly matched layer methods, and Dirichlet-to-Neumann approach. We analyze the query complexity of these algorithms, and perform numerical experiments to demonstrate the validity of this approach. This helps to bridge the gap between available quantum algorithms and computational models for quantum dynamics in unbounded domains.
\end{abstract}

\section{Introduction}

Quantum computing is an emerging technology that harnesses the laws of quantum mechanics to deliver unprecedented computational power~\cite{hidary2019quantum,nielsen2002quantum,preskill2018quantum,rieffel2011quantum}. Quantum algorithms operate on an $n$-qubit Hilbert space with dimension $2^n$, offering vast multidimensional spaces for computational models. Hence, it has a unique capability to handle large-scale scientific computing problems. A natural application is the PDEs from time-dependent Schrödinger equations (TDSE), which follow unitary evolutions and hence the wave functions can be coherently represented on quantum computers. Known as Hamiltonian simulations, a variety of efficient algorithms have been developed toward this end \cite{BCC15,BCK15,BCC17,LC17,LC19,CGJ19,kieferova2019simulating,an2021time,jin2022quantum,an2022time,fang2023time}.

While the TDSE is formulated in the entire physical space,
in practice, the computation has to be done in a bounded domain, typically in locations where the electron density is concentrated. For situations when the electrons are being emitted outside the computational domain, such as the photoionization process, or when they are being drawn away, such as the ionization process, this approach, however, can result in an extraordinarily large computational domain.  An artificial boundary condition (ABC) that may absorb outgoing wave packets and, more importantly, keep the size of the computational domain to a minimum, is frequently used to solve such problems.  A correctly set up ABC will yield the same result as if the computation is done in the infinite domain, e.g., see the survey paper \cite{antoine2008review}.

Due to the introduction of the ABCs, the computer simulation is now following a dynamics that is no longer unitary. As a result, existing Hamiltonian simulation techniques can not be directly applied. The goal of this paper is to bridge this emerging gap, by mapping the non-Hamiltonian dynamics back to a Schrödinger equation or Hamiltonian system, to enable the immediate
Hamiltonian simulation capability. This is accomplished by using the recently introduced Schrödingerization technique \cite{jin2022quantumshort, jin2022quantum}.  It is a generic framework to convert {\it any} linear partial (and ordinary) differential equation or dynamical system into a system of Schr\"odinger equations or a Hamiltonian system by going to the  space which is one dimensional higher, via the warped phase transformation. A key advantage of the Schr\"odingerization approach is its simplicity and generality, where the general Hamiltonian the system evolves under has a simple form. When the boundary value problem is considered, resulting in an inhomogeneous system that is not Hamiltonian, one just needs to introduce one auxiliary variable to make the system homogeneous, and then the Schr\"odingerization process still applies \cite{jin2022quantumshort, jin2022quantum}.

In this paper, we explore three popular types of ABCs, including the complex absorbing potential method, perfect-matched layers method, and Dirichlet-to-Neumann approach, to captivate the quantum implementation utilizing the Schr\"odingerization approach.

Another potential alternative to implementing the ABCs is by using quantum linear solvers (QLS) \cite{harrow2009quantum,childs2017quantum}.  This can be done by first performing spatial discretization that reduces the models to a system of linear ODEs, followed by time-marching schemes. Runge-Kutta methods, spectral (collocation) methods \cite{childs2019quantum}, multi-step methods \cite{berry2017quantum}, and Dyson series \cite{krovi2022improved}, are some of the important existing methods. In some earlier works \cite{childs2019quantum}, the analysis for this type of algorithm assumes that the matrix $A$ in the linear ODE system is diagonalizable, and the final complexity involves a condition number of the eigenvector matrix. This is later improved by Krovi  \cite{krovi2022improved}, where  the condition number  is replaced by the following bound,
  \[C(A) = \sup_{t\in [0,T]} \norm{\exp(At)}.\]
In the context of ABCs, due to the dissipative nature, the matrix $A$ is typically stable, in that the real parts of the eigenvalues are non-positive. Indeed, such stability property is a key focus in the development and analysis of ABCs \cite{engquist1979radiation}. The stability implies that $C(A)=1$, thus completely removing the dependence of the complexity on the condition number.

However, QLS-based methods still require more involved steps like a truncation of Taylor series expansions or the construction of a unitary operation associated with the inverse of a matrix. The latter is often realised through the block-encoding formalism \cite{gilyen2019quantum}. While these results might give similar query and gate complexity to the Schr\"odingerization approach, it is not straightforward to actually show explicitly how the inverse is obtained in the block-encoding formalism, which makes it difficult to perform in practice. This is in contrast to the Schr\"odingerization method, which is a much simpler scheme where the Hamiltonian to be simulated is {\it explicitly given}, which places the problem directly in the realm of Hamiltonian simulation. The cost in the Schr\"odingerization method is also {\it independent of the condition number}.

We note that there are also very recent alternatives to QLS for simulating amplitude-encoded solutions of differential equations, for instance schemes based on block-encoding \cite{an2022theory} or linear combination of unitaries (LCU) \cite{an2023linear}. The application of these methods to ABCs could provide a complementary viewpoint to our methods here.

\section{A brief review of existing boundary conditions}

\begin{figure}[!htb]
    \centering
    \includegraphics[scale=0.5]{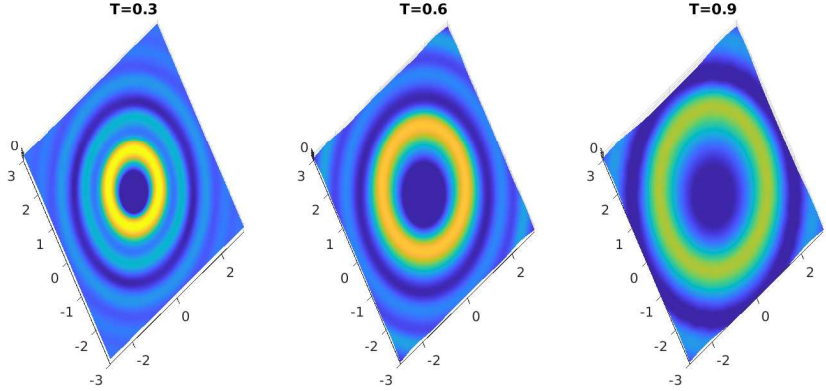}
    \caption{Waves propagating outside the computational domain $\Omega=[-3,3]\times [-3,3]$.}
    \label{fig:my_label}
\end{figure}

We first review several types of artificial boundary conditions that are commonly used in practice.
The goal of these numerical techniques is to be able to simulate the dynamics in the entire space $\mathbb{R}^d$, i.e.,
\begin{equation}  \label{eq:tdse}
  i\frac{\partial }{\partial t}\psi(\bm x, t) = \hat{H}\psi(\bm x, t), \quad \bm x \in \mathbb{R}^d,
\end{equation}
where $\hat{H}$ is a Hamiltonian operator expressed in reduced units,
\begin{equation}
    \hat{H} = - \frac{\nabla^2}{2} + V(\bm x,t).
\end{equation}
We assume that the initial wave function $\psi(\bm x, 0)$ is compactly supported in a subdomain,
 here denoted by $\Omega \subset  \mathbb{R}^d$. On the other hand, due to the initial momentum, or the influence of the external potential, the wave function can spread well beyond $\Omega$, as shown in Figure \ref{fig:my_label}.   To avoid the need for continuously expanding the computational domain, it is necessary to develop suitable boundary conditions that can propagate wave packets beyond the domain's boundaries. These conditions, known as artificial boundary conditions (ABCs), are intended to prevent boundary reflections, and hence the interference with the dynamics inside the domain.

\subsection{Complex Absorbing Potentials}

The idea of the complex absorbing potential (CAP) is to introduce an artificial potential with negative imaginary part to the TDSE. This modifies the TDSE \eqref{eq:tdse} to,
\begin{equation}
  \label{cap}
  i\frac{\partial }{\partial t}\psi(\bm x, t) = \hat{H}\psi(\bm x, t) +  W (\bm x)  \psi(\bm x, t), \quad \bm x \in D.
  \end{equation}

\begin{figure}[!htb]
    \centering
    \includegraphics[scale=0.5]{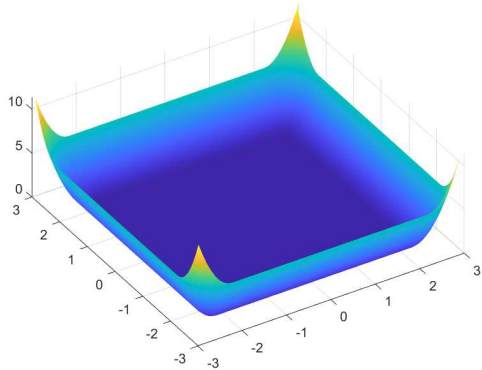}
    \caption{An example of the complex absorbing potential $W(x,y)$ for a two-dimensional domain { (The amplitude is for $-\text{Im} W$)}. $\Omega=[-2.2,2.2]\times [-2.2,2.2]$ and
    $D=[-3,3]\times [-3,3]$.
    Motivated by the study in \cite{riss1996investigation}, we choose a polynomial form $ W(x,y)= -10 i (\abs{x}-2.2)_+^2 - 10 i (\abs{y}-2.2)_+^2 $.  }
    \label{fig:sigma}
\end{figure}

Here $D \supset \Omega $ is a bigger domain that offers an absorbing layer, $D\backslash \Omega$, surrounding the computational domain $\Omega$.
The complex potential $W(\bm x)$  is selected so that it is zero in the computational domain $\Omega$ to keep the original dynamics unaltered, while in a surrounding absorbing layer, it has a negative imaginary part with magnitude slowly increased away from the boundary $\partial \Omega$ (see Figure \ref{fig:sigma} for example). The hope is that the wave function at the boundary of the absorbing layer is sufficiently damped at which point it can be simply set to zero. As a popular technique   in computational chemistry,  a wide variety of parametric forms for the imaginary potential have been proposed and extensively  tested \cite{muga2004complex,yu2018optimized}. Some of the choices may include a non-zero real part in $W(\bm x)$ to offer more flexibility.

From the computational perspective, the CAP method only introduces a modification to the diagonal elements of $\hat{H}$. The implementation is quite straightforward and the boundary conditions outside the absorbing layer can be chosen as the homogeneous Dirichlet boundary conditions.

\subsection{Perfectly Matched Layers methods}
Another classical strategy is the perfectly matched layer (PML) \cite{berenger1994perfectly}, where one first constructs a buffer layer so that the outgoing waves in the computational domain are exactly preserved (perfect matching). Although conceptually similar to CAP, the PML, at least at the continuous level, can theoretically absorb the outgoing waves. The most common mathematical approach to ensure the perfectly matching property  is to introduce a complex stretching of the spatial coordinate to derive a modified equation in the buffer layer, and then the resulting models are discretized simultaneously in the implementation.

PML has applications in many areas of science and engineering, such as electromagnetics, acoustics, and fluid dynamics, where accurate simulation of wave propagation is critical for understanding and predicting physical phenomena. For the TDSE \eqref{eq:tdse}, the PML  has been developed by Zheng \cite{zheng2007perfectly} and Nissen and Kreiss \cite{nissen2011optimized}.
We follow the derivation in \cite{nissen2011optimized}, and to briefly demonstrate the formulation, we first consider the 1d Schr\"{o}dinger equation,
\begin{equation}
    i \partial_t \psi = -\frac{1}{2}\partial_x^2 \psi + V(x).
\end{equation}

A derivation of a PML starts by replacing the time derivative with multiplication by $- i \omega$ with $\omega$ being the frequency variable. This changes the equation to,
\begin{equation}\label{eq: omega-psi}
     \omega \psi = -\frac{1}{2}\partial_x^2 \psi + V(x).
\end{equation}
In the second step, one performs a coordinate stretching,
\[ \partial_x \longrightarrow \frac{1}{1+ i\sigma(x)/\omega } \partial_x, \]
which reduces \eqref{eq: omega-psi} to
\[
\begin{aligned}
\omega \psi + i\sigma(x) \psi =& - \frac12 \partial_x  \left(\frac{1}{1+ i\sigma(x)/\omega } \partial_x \psi \right) + \big(1+ i\sigma(x)/\omega) V(x), \\
  = & -\frac{\partial_x^2}{2} \psi + V(x)  + \frac12 \partial_x  \left(\frac{i\sigma(x)/\omega}{1+ i\sigma(x)/\omega } \partial_x \psi \right)  + \frac{i\sigma(x)}{\omega} V(x).
\end{aligned}
\]

The selection of the function $\sigma$ may follow a similar recipe in the CAP method. Since we assume that $V=0$ outside the domain $\Omega$, we can drop the last term. We continue by defining an auxiliary function,
\[\chi = \frac{\sigma(x)/\omega}{1+ i\sigma(x)/\omega } \partial_x \psi.\]

Finally, by converting $- i \omega$ back to the time derivative, we obtain a system of equations,
\begin{equation}
    \begin{aligned}
        i\partial_t \psi = &\hat{H} \psi - i \sigma(x) \psi + \frac{i}{2} \partial_x \chi,\\
        i\partial_t \chi = & \sigma(x) \partial_x \psi - i \sigma(x) \chi.
    \end{aligned}
\end{equation}

We should point out that this idea of stretching the coordinate shows a close resemblance to the exterior coordinate scaling method \cite{sidky2000boundary,he2007absorbing}. But the corresponding modified equations  are quite different from the equations that we derived here.

For a two-dimensional problem, one  can repeat the same steps in the second coordinate and generalize the model as,
\begin{equation}\label{PML}
\left\{
    \begin{aligned}
        i\partial_t \psi = &\hat{H} \psi - i \sigma(x) \psi -i \sigma(y) \psi + \frac{i}{2} \partial_x \chi +  \frac{i}{2} \partial_y \phi,\\
        i\partial_t \chi = & \sigma(x) \partial_x \psi - i \sigma(x) \chi,\\
        i\partial_t \phi = & \sigma(y) \partial_y \psi - i \sigma(y) \phi,
    \end{aligned}\right. \quad (x,y)\in D.
\end{equation}

In the interior of $\Omega$, we have $\sigma \equiv 0$, and thus the new equations \eqref{PML} will be reduced to the TDSE \eqref{eq:tdse}. Meanwhile, outside the buffer layer, i.e., at $\partial D,$ the wave function can be set to zero. Notice that the perfectly matching property no longer holds after the numerical discretization. Therefore, a suitable choice of $\sigma$, together with a sufficiently large buffer layer, is crucial to the performance of a PML.

\subsection{The Dirichlet-to-Neumann Approach}

The third type of ABC is based on a Dirichlet-to-Neumann (DtN) map. One can first apply a semi-discrete scheme to the TDSE \eqref{eq:tdse}, e.g., a standard finite difference scheme for the kinetic energy. This still yields an infinite system of ODEs.  The next step is a domain decomposition. Toward this end,  we define $\Omega_\i$ to be the grid points in $\Omega$, and similarly, let $\Omega_\ii$ be the set of grid points outside the domain $\Omega.$

The semi-discrete model can be written in the following compact form:
\begin{equation}
  \label{eq:22}
  \left\{
    \begin{aligned}
      i\dot{\bm \psi}_\i(t) &= H_{\ici}(t)\bm \psi_\i(t)+H_{\icii}(t)\bm \psi_\ii(t),  \\
      i\dot{\bm \psi}_\ii(t) &= H_{\iici}(t)\bm \psi_\i(t)+H_{\iicii}(t)\bm \psi_\ii(t),
    \end{aligned}
  \right.
\end{equation}
where $\bm \psi_\i = [ \psi(\bm x_k)]_{k\in \Omega_\i}$ and $\bm \psi_\ii= [ \psi(\bm x_k)]_{k\in \Omega_\ii}$. $\bm \psi_\i\in \mathbb C^{n_\i}$ and $\bm \psi_\ii\in \mathbb C^{n_\ii}$. We denote the discretization of the Hamiltonian operator in a partitioned form,
\begin{equation}
  \label{eq:32}
  H =
  \begin{bmatrix}
    H_\ici & H_\icii\\
    H_\iici & H_\iicii
  \end{bmatrix}.
\end{equation}
Since the Schr\"{o}dinger equation \eqref{eq:tdse} has been descretized in space,
we use $\dot{\quad}$ to denote the time derivatives hereafter.  In addition,  since we assumed $V=0$ in $\Omega_\ii$, the operator $H_\iicii$ only contains the kinetic energy (Laplacian).

For the grid points in the computational domain $\Omega_\i$, one can separate out the grid points next to the boundary. Denote the set of those grid points by $\Gamma$:
\[\Gamma = \{ \bm{x}_j \in \Omega_\i: \mbox{if there exists $\bm{x}_k \in \Omega_\ii$ such that $H_{jk} \ne 0$} \}.\]
By reordering the grid points, one can arrange $\bm \psi_\i$ as follows
\begin{equation}\label{eq: reorder}
\bm \psi_\i =
\begin{bmatrix}
  \bm \psi_\Gamma\\
  \bm \psi_{\i\backslash\Gamma}
\end{bmatrix}.
\end{equation}
As a result, the off-diagonal block of the Hamiltonian may be written as
\begin{equation*}\label{eq: h2g}
H_\iici = \big[H_{\iiciii} \;\; H_{\ii,\i\backslash\Gamma}\big]=\big[H_{\iiciii} \;\; \bm 0\big].
\end{equation*}

By using Laplace transform, Wu and Li (see Section II in \cite{wu2020absorbing}) derived the following exact boundary condition,
\begin{equation}\label{eq: dtn}
\left\{
\begin{aligned}
     \dot{\bm \psi}_\i(t) = & -iH_\ici\bm \psi_\i(t) -i E^T \bm \phi_\Gamma(t), \\
     \bm{\phi}_\Gamma(t) = & \int_0^t \kappa(t-\tau) \bm{\psi}_\Gamma(\tau) d \tau,
\end{aligned}\right.
\end{equation}

Here $E$ is a restriction operator to extract the components of a  wave function that correspond to grid points at the boundary $\Gamma$ from a function defined in  $\Omega_\i$, i.e.,
\begin{equation*}\label{eq: phig}
\bm \psi_\Gamma=E\bm \psi_\i.
\end{equation*}
Furthermore, $\bm \phi_\Gamma$ represents the influence of the wave functions in $\Omega_\ii$ on the wave functions in $\Omega_\i$,
\begin{equation*}
  \bm \phi_\Gamma=H_{\iiicii}\bm \psi_\ii(t).
\end{equation*}
But this connection is not needed in the implementation of \eqref{eq: dtn}.

The second equation in \eqref{eq: dtn} is regarded as discrete DtN map, where $\kappa(t)$ is the real-time kernel function which corresponds to $K(s)$ in the Laplace domain with
\[K(s) = - H_{\Gamma, \ii}(H_{\ii,\ii} -i sI)^{-1} H_{\ii, \Gamma}.\]
Such ABCs can also be derived without the spatial discretization  \cite{han2004exact,antoine2008review}. For example, for one-dimensional problems, the integral term can be expressed as a fractional derivative \cite{arnold1998numerically,antoine2008review}. But a discretization would be needed afterward.

Up to this point, the formulation is exact. In practice, $K(s)$ is often treated using a Pad\'e approximation. For instance, a zeroth order approximation involves selecting an $s_0$ as the interpolation point, and let
\begin{equation}\label{eq: Rmat}
R= K(s_0)= -H_{\iiicii} \left(H_\iicii - is_0 I\right)^{-1}H_{\iiciii}.
\end{equation}
Combining with \eqref{eq: dtn}, the ABC becomes,
\begin{equation}
  \label{eq:dynamics0}
  \dot{\bm{\psi}}_\i(t)= -iH_\ici\bm{\psi}_\i(t) - iE^T R E \bm \psi_\i(t).
\end{equation}
This approximation also adds a potential with the anti-Hermitian part being negative definite \cite{wu2020absorbing}. Thus it can be considered as a CAP method; But the Hermitian part is not necessarily zero and the anti-Hermitian part is usually not diagonal.

From the first-order Padé approximation, one obtains the ABC as follows,
\begin{equation}
  \label{eq:dynamics1}
  \left\{
    \begin{aligned}
      \frac{d}{d t}\bm \psi_\i(t) &= -iH_\ici\bm \psi_\i(t) - iE^T\bm \phi_\Gamma(t),\\
      \frac{d}{d t}\bm \phi_\Gamma(t) &= B\bm \phi_\Gamma(t) + A E \bm \psi_\i(t).
    \end{aligned}
  \right.
\end{equation}
By a straightforward interpolation at $s=s_0$, one has (see the proof of Theorem 3.2 in \cite{wu2020absorbing}),
\begin{equation}\label{DtNm1}
A =  -i H_{\iiicii} H_{\iiciii}, \qquad B =  s_0 I - A R^{-1}.
\end{equation}

We have presented the DtN approach as if $H_\iicii$ is a finite matrix.  One can take an infinite volume limit by imposing a zero far-field condition.  Another important observation is that  the matrix $R$ in \eqref{eq: Rmat} involves the operator $H_\iicii$ in the exterior, but the DtN map can be computed from a discrete boundary element equation \cite{martinsson2009boundary,li2012atomistic}, which only involves the wave function at the boundary. Such equations can be further simplified by seeking a sparse approximation of $R$ using least-squares \cite{li2009efficient}.

\section{Quantum simulations via Schr\"odingerization}

From the proceeding section, we observe that the TDSE under the three types of ABCs can be expressed in the following general form,
\begin{equation}\label{TDSEgeneral}
    \frac{d}{dt} \ket{\phi(t)}= -i H_0  \ket{\phi(t)}  - H_1  \ket{\phi(t)}.
\end{equation}
Here $\ket{\phi(t)}$ is a quantum state that encodes the wave function, possibly the auxiliary functions,  in the subdomain. $H_0$ and $H_1$ are both Hermitian, and $H_0$ is a slight modification of the original Hamiltonian.

The  Schr\"odingerization approach was first proposed in \cite{jin2022quantum}. The key step is the warped phase transformation $\bm{v}(t,p) = e^{-p} \bm{\phi}(t)$, defined for $p>0$ and symmetrically extends the initial data to $p<0$. This transformation converts \eqref{TDSEgeneral} to a system of linear convection equations:
\begin{equation}\label{u2v}
\begin{cases}
 \frac{d}{d t} \bm{v}(t,p) = -i H_0  \bm{v} + H_1 \partial_p \bm{v}, \\
 \bm{v}(0,p) = e^{-|p|} \bm{\psi}_0.
 \end{cases}
\end{equation}
On the other hand, the solution $\bm{\phi}(t)$ can be restored by a simple integration,
\begin{equation}\label{v2psi}
    \bm{\psi}(t) = \int_0^\infty \bm{v}(t,p) d p.
\end{equation}

A more intuitive view is by discretizing the $p$ dimension and concatenating the corresponding function for each $p$. Toward this end,  we observe that $\bm{v}(t,p)$ has an exponential decay in the $p$ space. Hence we may choose a sufficiently large interval, such that,
\[
\int_0^\infty \bm{v}(t,p) d p = \int_a^b \bm{v}(t,p) d p + \mathcal{O}(\epsilon).
\]
Next, we choose sub-intervals with size $\Delta p = (b-a)/M$  with $M=2N$ being an even number and the grid points denoted by $a = p_0<p_1<\cdots<p_M = b$. Then the mapping from $\bm v$ to $\psi$ can be implemented using a quadrature formula,
\[
\int_a^b \bm{v}(t,p) d p  \approx  \frac{\Delta p}{2} \left[ \bm{v}(t,a) + \bm{v}(t,b) \right]
 + \Delta p \sum_{m=1}^{M-1} \bm{v}(t,p_m).
\]

To compute $\bm v(t,p),$ let the vector $\bm{w}$ be the collection of the function $\bm v$ at these grid points, defined more precisely as follows,
\[\bm{w} = [\bm{w}_1; \bm{w}_2; \cdots; \bm{w}_n], \]
with ``;'' indicating the straightening of $\{\bm{w}_i\}_{i\ge 1}$ into a column vector. This can also be expressed as a superposition state using $\ket{k}$ as a new basis,
\[
 \bm{w}_i = \sum_k \bm{v}_i (t,p_k) \ket{k},
\]

By applying the discrete Fourier transformation in the $p$  direction, one arrives at,
\begin{equation}\label{heatww}
\frac{d}{d t} \bm{w}(t) = -i ( H_0 \otimes I ) \bm{w} + i (H_1 \otimes P_\mu) \bm{w}.
\end{equation}
 Here, $P_\mu$ is the matrix expression of the momentum operator $-i\partial_p$, given by
\[P_\mu = \Phi D_\mu \Phi^{-1},  \qquad D_\mu = \text{diag}(\mu_{-N}, \cdots, \mu_{N-1}), \]
where $\mu_l = 2\pi l/(b-a)$ are the Fourier modes and
\[\Phi = (\phi_{jl})_{M\times M} = (\phi_l(x_j))_{M\times M}, \qquad \phi_l(x) = e ^{i \mu_l (x-a)} \]
is the matrix representation of the discrete Fourier transform.

At this point, we have successfully mapped the dynamics back to a Hamiltonian system. By a change of variables $\tilde{\bm{w}} = (I_u \otimes \Phi^{-1})\bm{w}$, one has
\begin{equation}\label{generalSchr}
\frac{d}{d t} \tilde{\bm{w}}(t) = -i ( H_0 \otimes I) \tilde{\bm{w}} + i (H_1 \otimes D_\mu ) \tilde{\bm{w}}.
\end{equation}
This is more amenable to an approximation by a quantum algorithm. In particular, if $H_0$ and $H_1$ are sparse, then  \eqref{generalSchr} is a Schr\"odinger equation with the Hamiltonian
 $H = H_0 \otimes I - H_1 \otimes D_\mu$ that inherits the sparsity.

\subsection{Schr\"odingerization for the CAP method}

As an example, we consider the CAP method \eqref{cap} applied to the TDSE \eqref{eq:tdse} in two dimensions ($\bm x =(x,y)$), although the extension to higher dimensions is straightforward. As an example,
the five-point finite difference is used to discretize the Laplacian on the grid points,
\[x_0<x_1<\cdots<x_N<x_{N+1}, \qquad y_0<y_1<\cdots<y_N<y_{N+1}.\]

After a spatial discretization of \eqref{cap}, the semi-discrete system for the unknowns $\psi(x_i,y_j,t)$ for $1\le i, j \le N$  has the following form:
\begin{equation}\label{semiODEs}
\begin{cases}
\frac{d}{dt} \bm{\psi}(t) = -i H_0 \bm{\psi}(t) - H_1 \bm{\psi}(t), \\
\bm{\psi}(0) = \bm{\psi}_0,
\end{cases}
\end{equation}
where $H_0$ is a real symmetric matrix corresponding to the discretization of $\hat{H}$, together with the real parts of the absorbing potential $W(x,y)$, and  $ H_1 = -\text{diag}(\text{Im} \bm{W})$ is a diagonal matrix given by the imaginary part $W$ of the complex potential, where $\bm{W} = \sum_{ij} W(x_i,y_j) \ket{i,j}$.
In fact, let $D_{xx}$ be the one-dimensional difference matrices for the second-order derivatives ($\partial_{xx}$) under the homogeneous boundary conditions. Then
\begin{equation}\label{H01CAP}
H_0 = - \frac{1}{2}(D_{xx} \otimes I + I \otimes D_{xx}) + \bm{V} + \text{diag}(\text{Re} \bm{W}) , \qquad \bm{V} = \text{diag}(V(x_i,y_j)),
\end{equation}
where we have assumed the same partitions along $x$ and $y$ directions.

\subsection{Schr\"odingerization for the PML method}

The PML in \eqref{PML} involves both first and second-order derivatives.  By applying the central differences to these derivatives, we obtain a straightforward discretization,
\[
\begin{cases}
\partial_t \bm{\psi} = -i H_V \bm{\psi} - (I\otimes \bm{\sigma}_x + \bm{\sigma}_y \otimes I) \bm{\psi} + \frac{1}{2} (I\otimes D_x) \bm\chi + \frac{1}{2} (D_y \otimes I) \bm\phi,\\
\partial_t \bm{\chi} = -i (I \otimes (\bm{\sigma}_xD_x)) \bm\psi -  (I\otimes \bm\sigma_y) \bm\chi,\\
\partial_t \bm{\phi} = -i ( (\bm\sigma_y D_y) \otimes  I)\bm\psi -  (\bm\sigma_y \otimes I) \bm\phi,
\end{cases}
\]
where $D_x$ and $D_{xx}$ are the one-dimensional difference matrices for finite-difference approximations of the first- and second-order derivatives ($\partial_x$ and $\partial_{xx}$) under homogeneous boundary conditions, $\bm{\sigma}_x = \text{diag}(\sigma(x_1), \cdots, \sigma(x_N))$ is a diagonal matrix and
\begin{equation}\label{SchH}
H_V = - \frac{1}{2}(D_{xx} \otimes I + I \otimes D_{xx}) + \bm{V}, \qquad \bm{V} = \text{diag}(V(x_i,y_j)).
\end{equation}
Let $\bm{u}_h(t) = [\bm{\psi};  \bm{\chi}; \bm{\phi}]$. Then the above system can be written in matrix form,
\[\frac{d}{d t} \bm{u}_h = L_h \bm{u}_h,\]
where
\begin{equation}\label{PMLLh}
L_h = \begin{bmatrix}
 -i H_V - (I\otimes \bm{\sigma}_x + \bm{\sigma}_y \otimes I) & \frac{1}{2} (I\otimes D_x) &  \frac{1}{2} (D_y \otimes I) \\
 -i (I \otimes (\bm{\sigma}_xD_x))    &  -  (I\otimes \bm\sigma_x)  &  O \\
 -i ( (\bm\sigma_y D_y) \otimes  I)   &   O   &   -  (\bm\sigma_y\otimes I)
 \end{bmatrix}.
\end{equation}
In accordance with \eqref{semiODEs}, one has $H_0 = -(L_h - L_h^\dag)/(2i)$ and $H_1 = -(L_h + L_h^\dag)/2$.

\subsection{Schr\"odingerization for the DtN approach}

Before we apply the Schr\"odingerization procedure for the DtN method \eqref{eq:dynamics1}, we
first introduce a change of variables, by defining,
\begin{equation}
  \bm  \psi_\Gamma(t) = \left( H_{\iiicii} H_{\iiciii} \right)^{-1/2} \bm   \phi_\Gamma(t).
\end{equation}
The invertibility of the matrix can be seen from the linear independence of the rows of the matrix $H_{\iiicii}$.

Direct substitutions into \eqref{eq:dynamics1} yield,
\begin{equation}
  \label{eq:dynamics-AB}
  \left\{
    \begin{aligned}
      \frac{d}{d t}\bm \psi_\i(t) &= -iH_\ici\bm \psi_\i(t) - i \Sigma_{\i,\Gamma} \bm \psi_\Gamma(t),\\
      \frac{d}{d t}\bm \psi_\Gamma(t) &= -i \Sigma_{\i,\Gamma}^T \bm \psi_\i(t) + \Sigma_{\Gamma,\Gamma}  \bm \psi_\Gamma(t).
    \end{aligned}
  \right.
\end{equation}
Here  we have defined matrices,
\begin{equation}\label{self-e}
  \begin{aligned}
         S = &  H_{\iiicii} H_{\iiciii}, \\
        \Sigma_{\i,\Gamma} & = E^T S^{1/2}, \\
        \Sigma_{\Gamma,\Gamma} & = s_0 I +i S^{1/2} R^{-1} S^{1/2}.
  \end{aligned}
\end{equation}
These notations are motivated by the notions of self-energy in electron transport \cite{brandbyge2002density}.

More explicit expressions for these matrices can be obtained using the formula \eqref{eq: Rmat}:
\[
\begin{aligned}
    R =& -H_{\iiicii} \left(H_\iicii - is_0 I\right)^{-1}H_{\iiciii}, \\
     = & -H_{\iiicii} \left(H_\iicii^2 + s_0^2 I\right)^{-1}\left(H_\iicii + is_0 I\right) H_{\iiciii} \\
     = & - G_0 - i s_0 G_1, \\
\end{aligned}
\]

Here to arrive at the second line, we have used the trivial fact that $H_\iicii$ commutes with $I$. In the third line, we have defined the matrices
\begin{equation}\label{G0G1}
 G_0=  H_{\iiicii} \left(H_\iicii^2 + s_0^2 I\right)^{-1} H_\iicii
H_{\iiciii}, \quad  G_1=  H_{\iiicii} \left(H_\iicii^2 + s_0^2 I\right)^{-1}
H_{\iiciii}.
\end{equation}

To separate out the Hermitian and skew-Hermitian parts of $\Sigma_{\Gamma,\Gamma}$, we notice that,
\[
 \frac12 \left( R^{-1} + \left( R^{-1}\right)^\dagger \right)
= \frac12  R^{-1} \left( R^{\dagger} + R \right) \left( R^{-1}\right)^\dagger= -R^{-1} G_0 \left( R^{-1}\right)^\dagger.
\]
Similarly, we have,
\[
 \frac1{2i} \left( R^{-1} - \left( R^{-1}\right)^\dagger \right)
= \frac1{2i}  R^{-1} \left( R^{\dagger} - R \right) \left( R^{-1}\right)^\dagger= -R^{-1} G_1 \left( R^{-1}\right)^\dagger.
\]

Now we can apply
Schr\"odingerization procedure
with  the Hermitian and skew-Hermitian parts given by,
\begin{equation}\label{H0-H1-dtn}
    H_0= \left[
    \begin{array}{cc}
       H_\ici  & \Sigma_{\i,\Gamma} \\
        \Sigma_{\i,\Gamma}^T  &-S^{\frac12} R^{-1} G_0 \left( R^{-1}\right)^\dagger S^{\frac12}
    \end{array} \right], \quad
    H_1= \left[
    \begin{array}{cc}
       \bm 0  & \bm 0  \\
       \bm 0   & -s_0I + S^{\frac12} R^{-1} G_1 \left( R^{-1}\right)^\dagger S^{\frac12}
    \end{array} \right].
\end{equation}

\section{Numerical Results}
We now present results from some numerical experiments.
As a concrete example, we consider the TDSE in two dimensions, and choose a square domain $\Omega:= [-3,3] \times [-3,3].$ Following the example in \cite{zheng2007perfectly}, we choose the initial condition as follows,
\begin{equation*}
\psi_0(x,y)=
    \begin{cases}
        1 + \cos (\pi r) + i (\cos (2\pi r) -1),  &    r \leq 1,\\
         0 &  \text{otherwise}.
    \end{cases}
\end{equation*}
Here $r= \sqrt{x^2+y^2}.$ For the potential, we choose $V(x,y)=\sin (2\pi r)$ in the unit disc and extend it to zero outside.

In the numerical experiments, we  solve \eqref{generalSchr} using the matrix exponential:
\[\tilde{\bm{w}}(t) = e^{-i H t} \tilde{\bm{w}}(0).\]
This is implemented on a classical computer using MATLAB's  built-in function {\it expm.m}.  In the implementation, we also take $N_x = N_p = 64$ and $p \in [L, R] = [-5, 5]$.  It is important, however, to note that the purpose of these experiments is to demonstrate that the equation \eqref{generalSchr} in the Schr\"odingerization captures the dynamics under the ABCs. A quantum implementation of \eqref{generalSchr} is much more feasible than classical computers, due to the less dependence on the dimension.

\begin{figure}[!htb]
  \centering
   \subfigure[$T = 0.3$]{\includegraphics[scale=0.34]{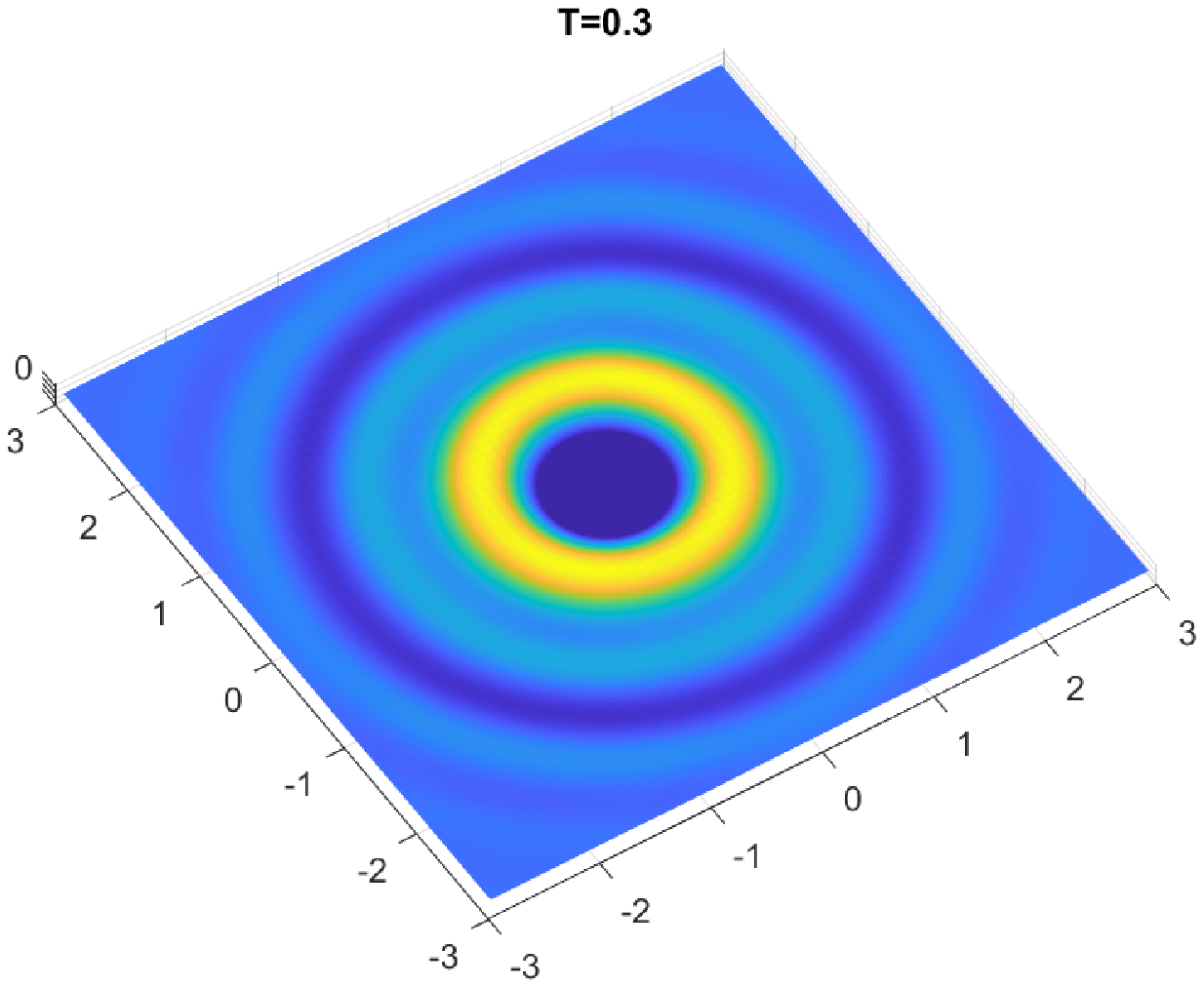}}
  \subfigure[$T = 0.6$]{\includegraphics[scale=0.34]{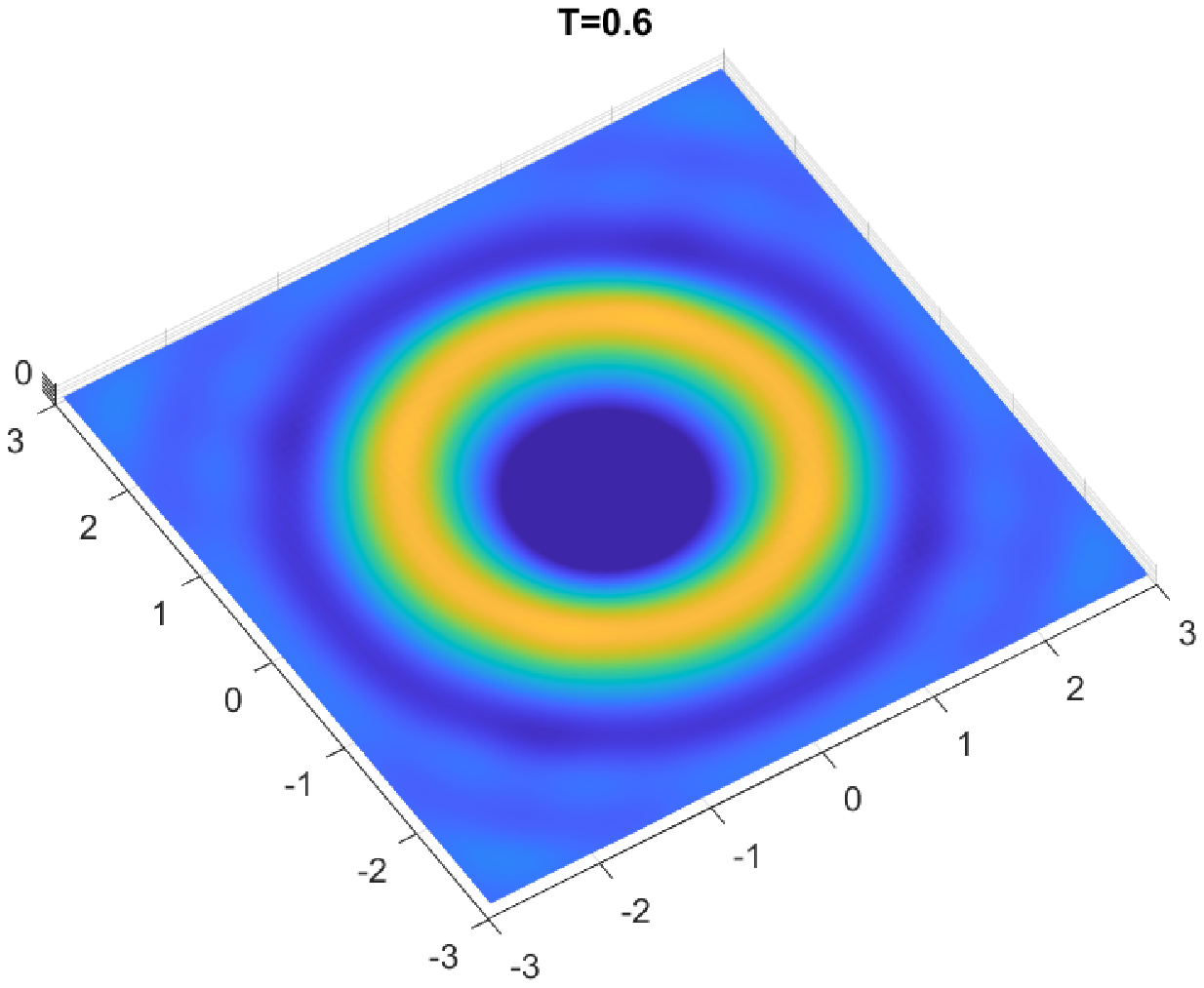}}
  \subfigure[$T = 0.9$]{\includegraphics[scale=0.34]{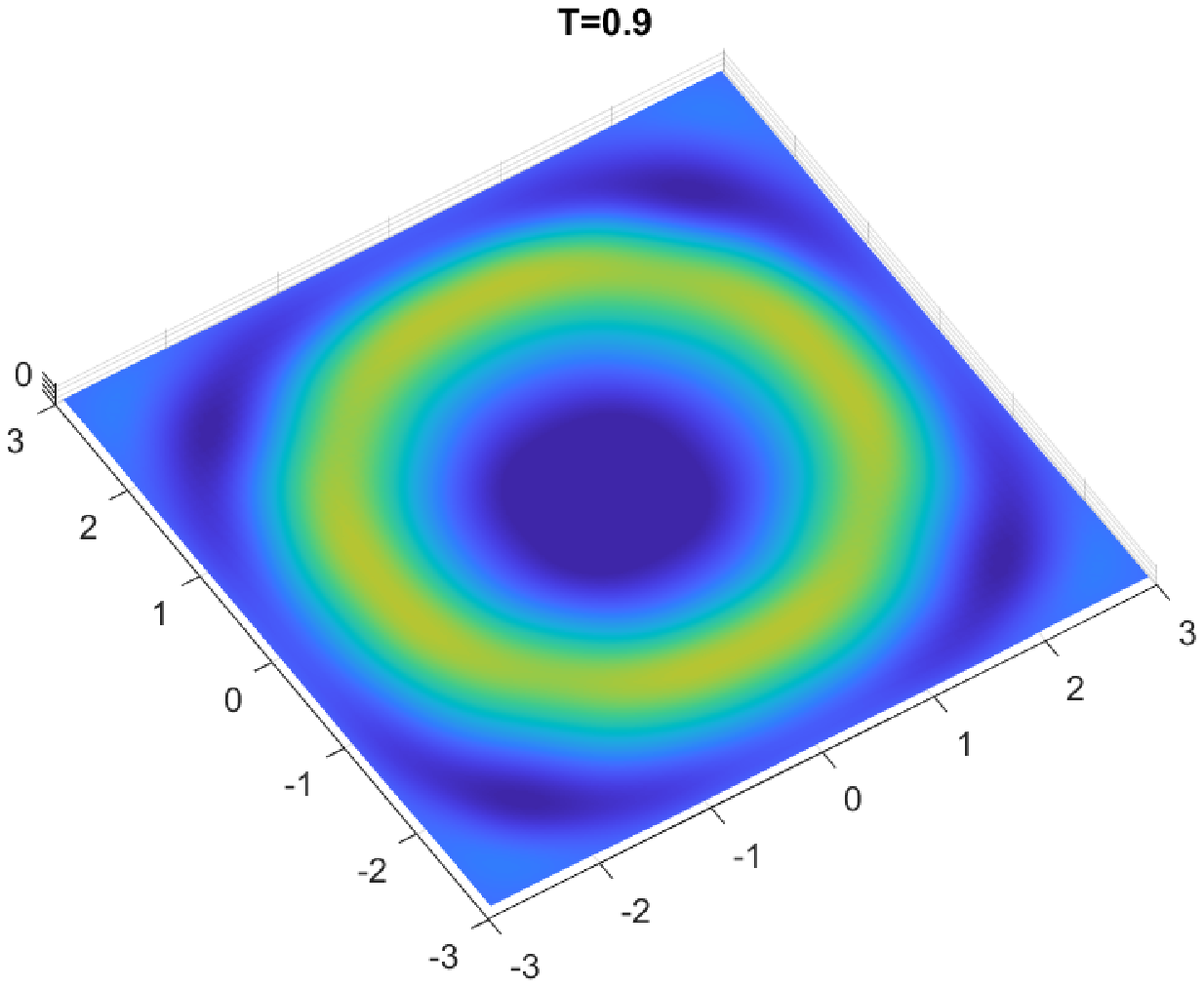}}\\
  \subfigure[Schr\"odingerization: $T = 0.3$]{\includegraphics[scale=0.34]{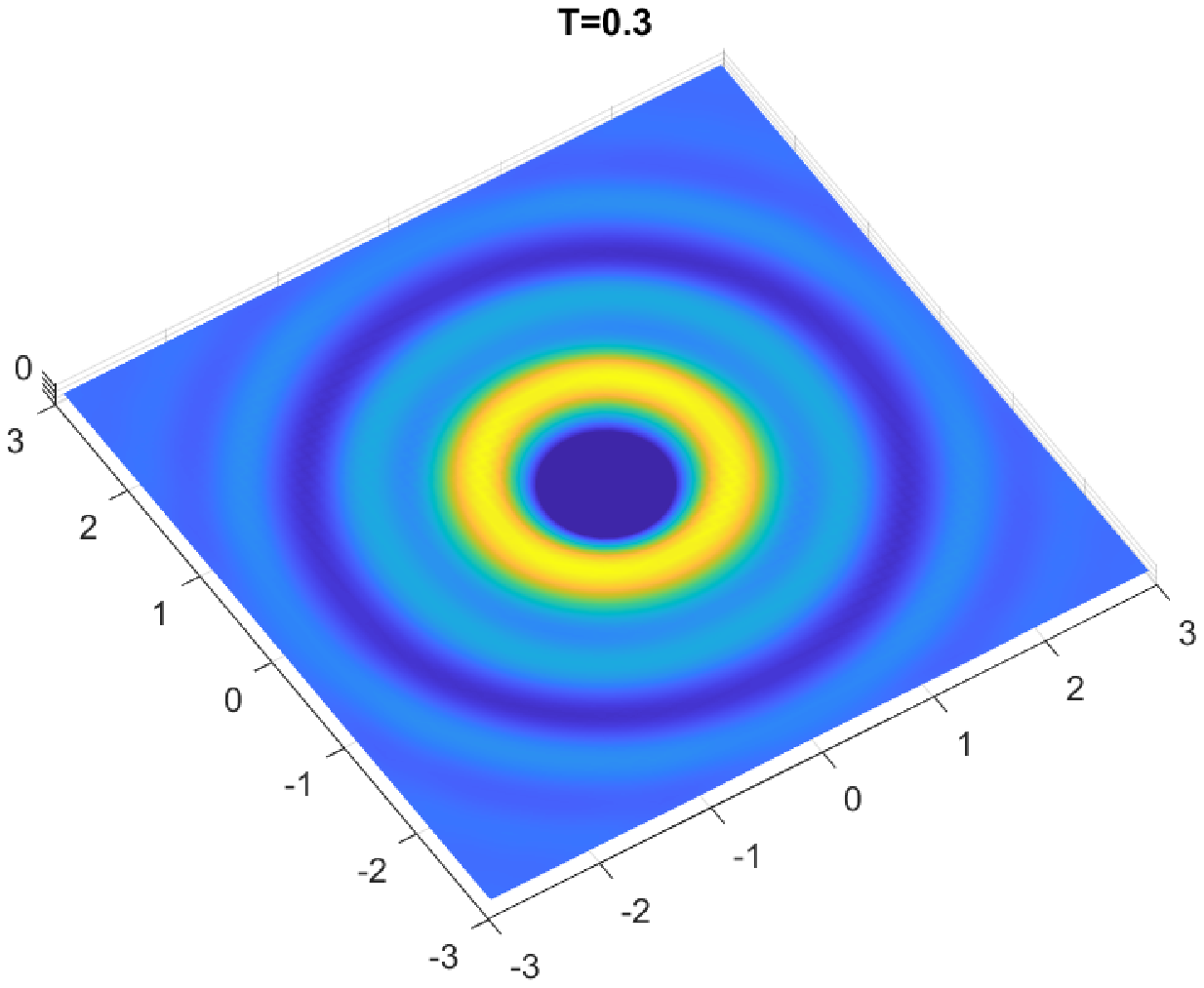}}
  \subfigure[Schr\"odingerization: $T = 0.6$]{\includegraphics[scale=0.34]{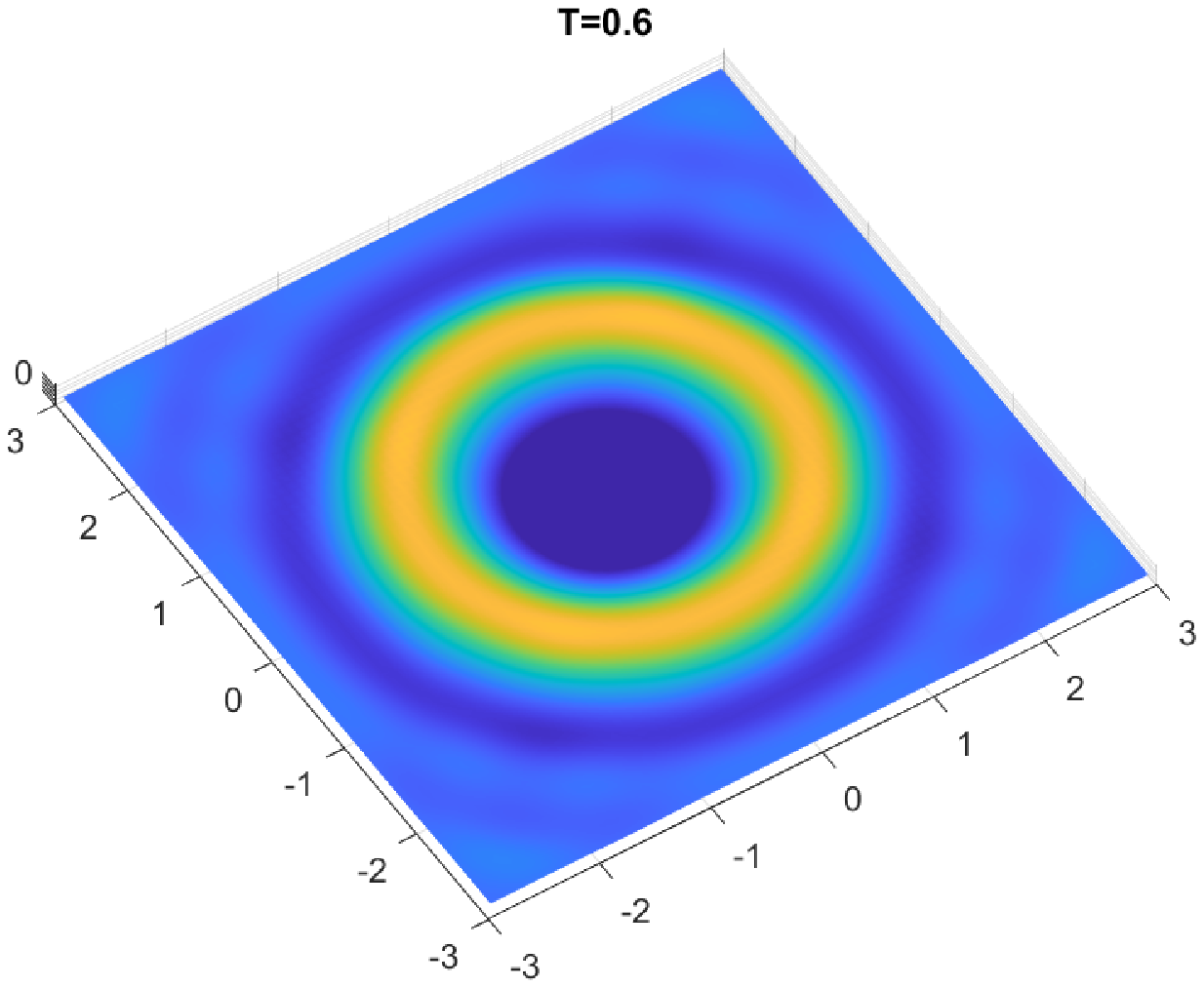}}
  \subfigure[Schr\"odingerization: $T = 0.9$]{\includegraphics[scale=0.34]{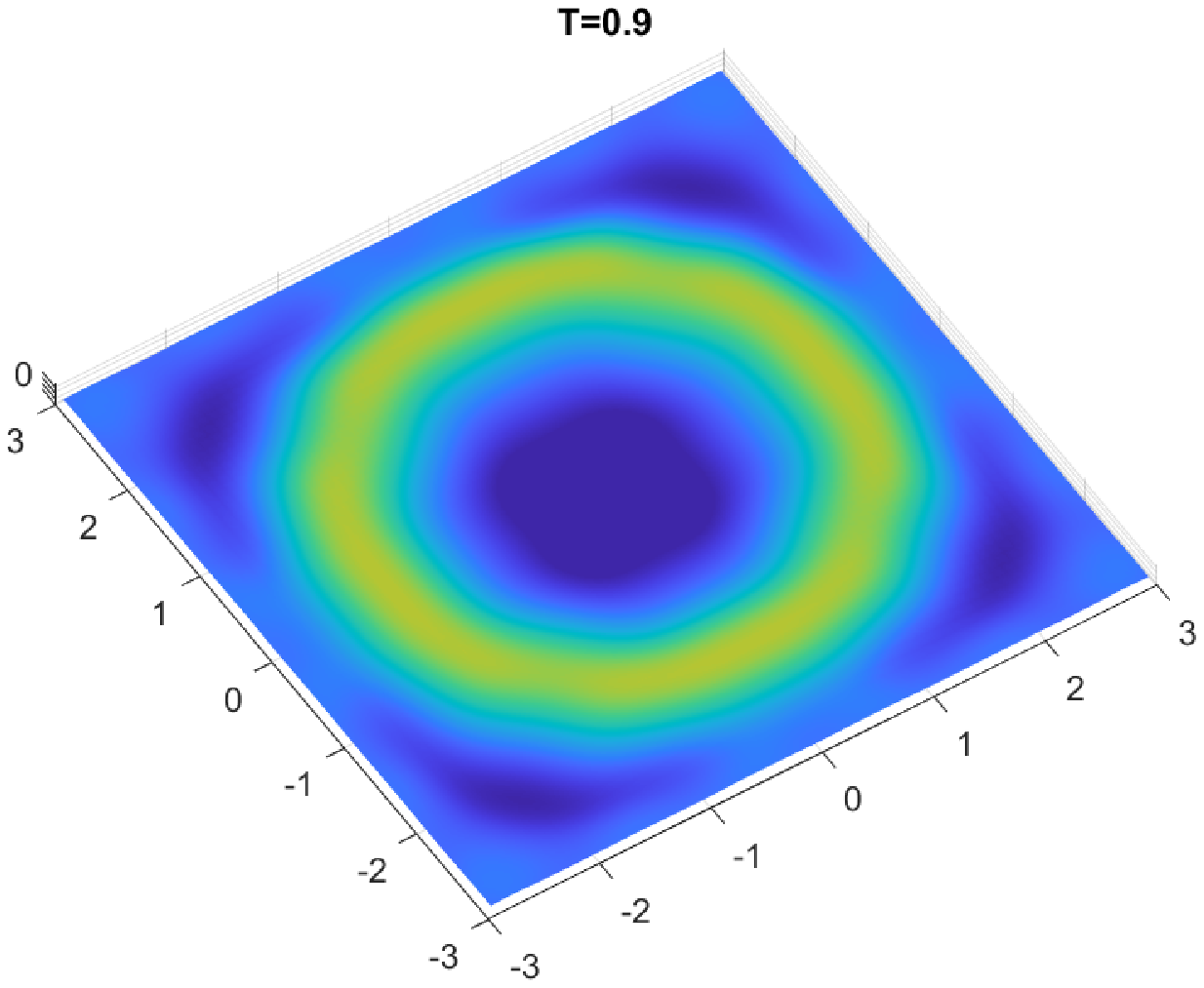}}\\
  \caption{Snapshots of the real parts of $\psi$ computed using the CAP method. Top: a direct implementation of \eqref{cap};  Implementation of the Schr\"odingerization form \eqref{generalSchr}.}\label{fig:SchrT0306}
\end{figure}

We first examine the CAP method. In these experiments, the imaginary potential $W(x,y)$ is chosen as the sum of $w_1(x,y)$ and $w_2(x,y)$, with
\[
w_1(x,y) =
    \begin{cases}
        - 10 i (|x|-2.2)^2,   &    |x| > 2.2\\
        0,   &    \mbox{otherwise}
    \end{cases}, \qquad
    w_2(x,y) =
    \begin{cases}
        - 10 i (|y|-2.2)^2,   &    |y| > 2.2\\
        0,   &    \mbox{otherwise}
    \end{cases},\]
as shown in Figure \ref{fig:sigma}.
 The results from quantum simulations of \eqref{cap} and \eqref{generalSchr} at $t = 0.3, 0.6$ and $t = 0.9$ are displayed in Figure \ref{fig:SchrT0306}(d-f), from which we observe very similar results to the ones given in Figure \ref{fig:SchrT0306}(a-c) for the original form \eqref{cap}.  This verifies the correctness of the protocol of the Schr\"odingerization approach.

\bigskip

\begin{figure}[!htb]
  \centering
  \subfigure[Original: $T = 0.3$]{\includegraphics[scale=0.34]{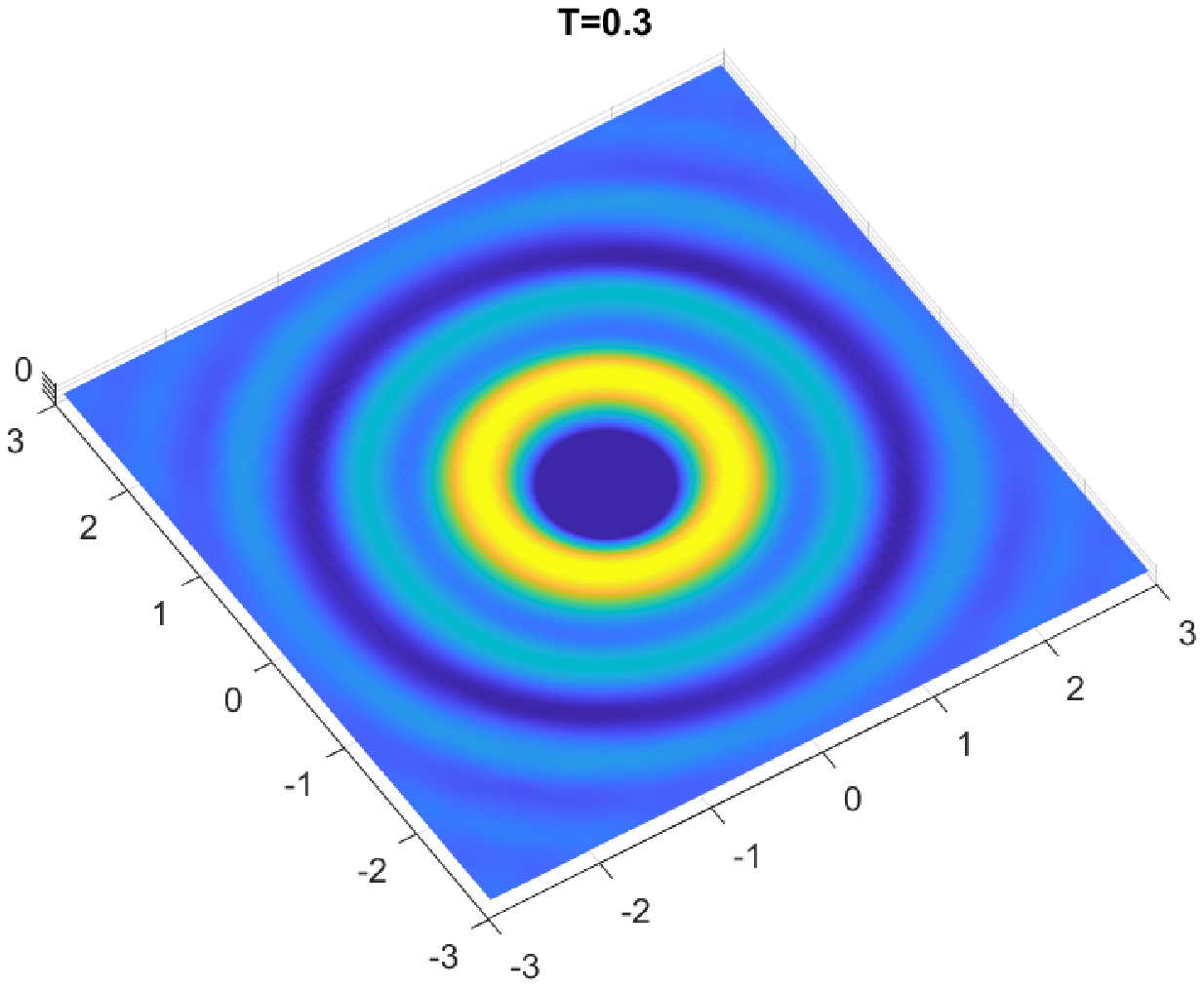}}
  \subfigure[Original: $T = 0.6$]{\includegraphics[scale=0.34]{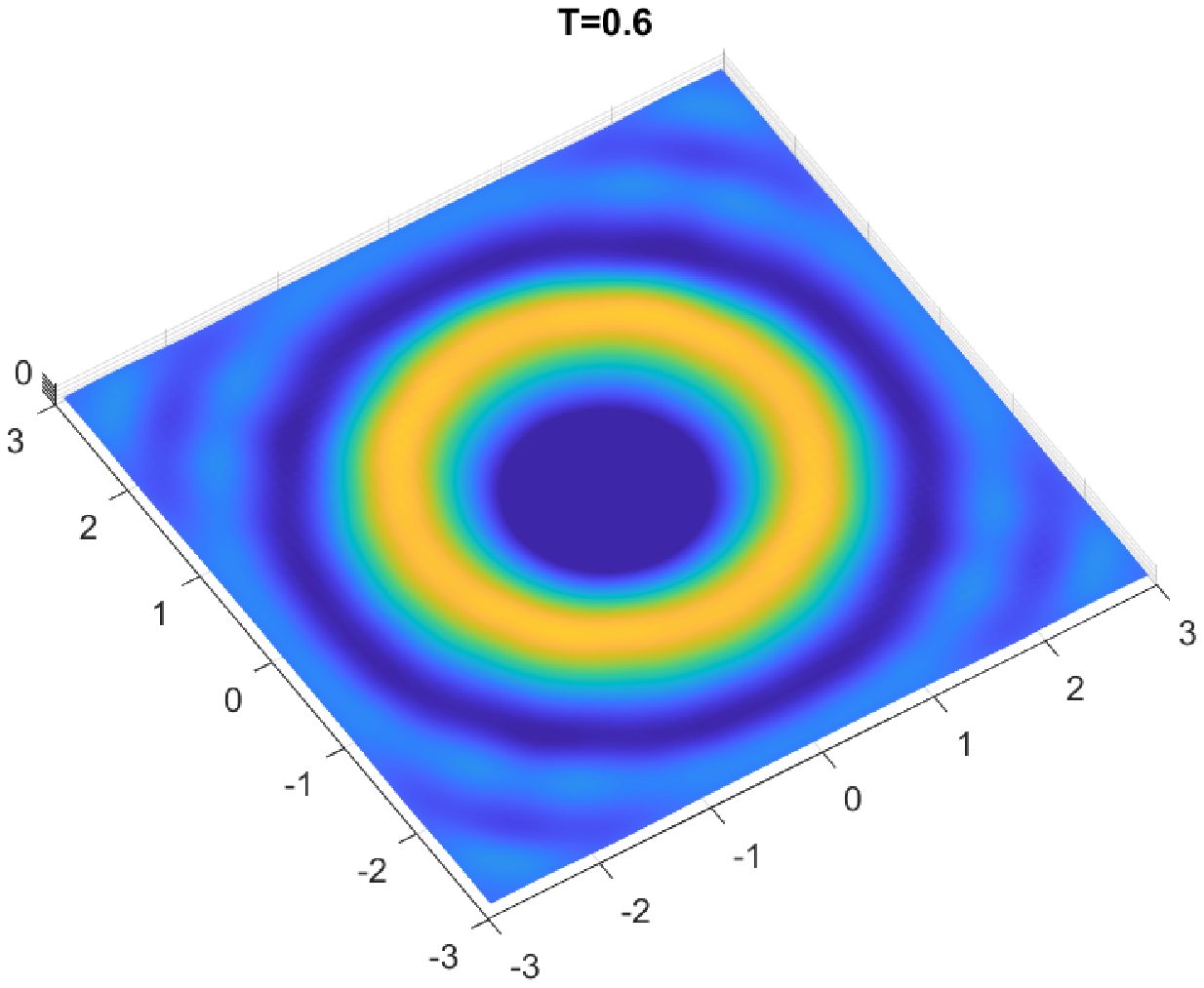}}
  \subfigure[Original: $T = 0.9$]{\includegraphics[scale=0.34]{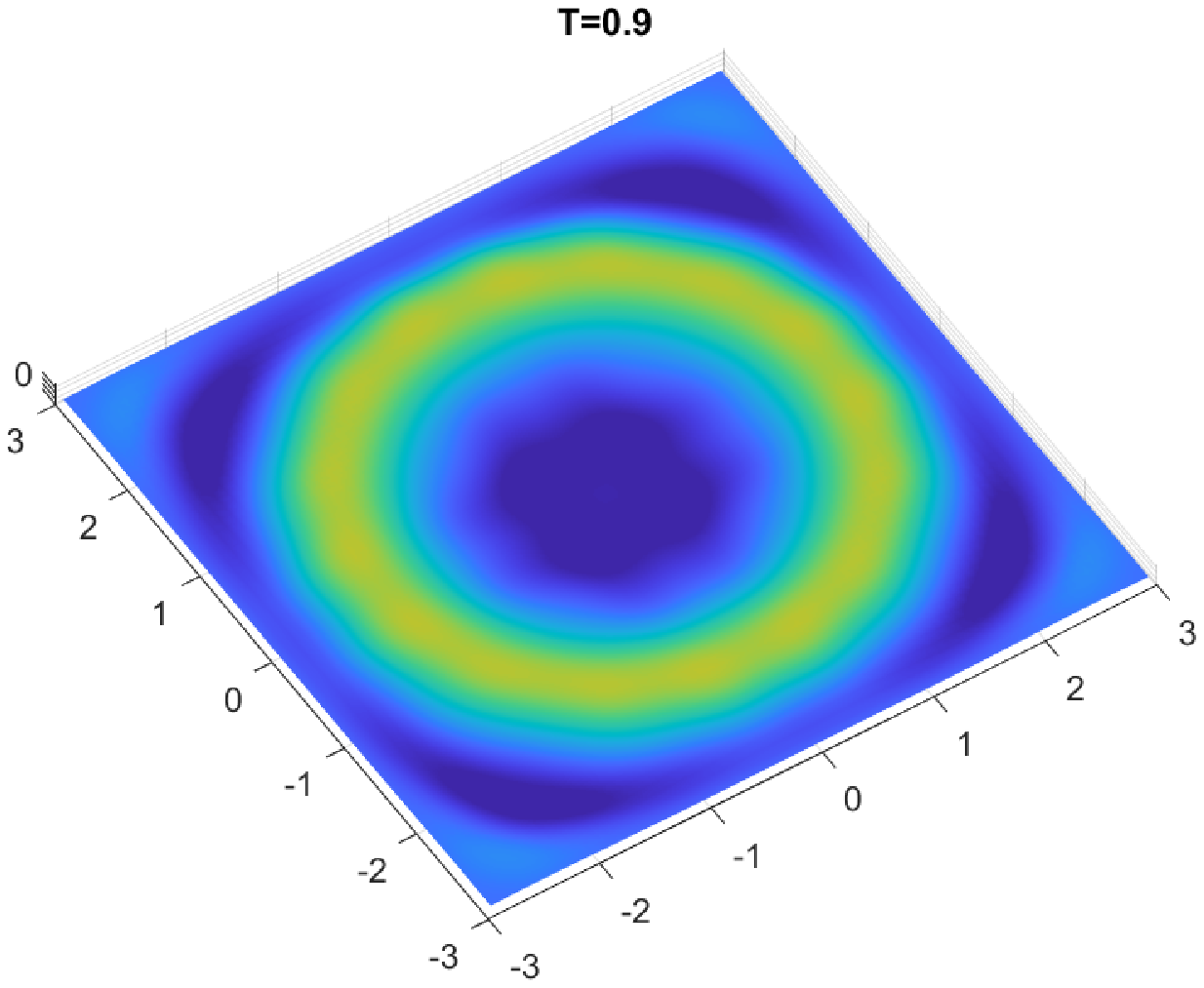}}\\
  \subfigure[Schr\"odingerization: $T = 0.3$]{\includegraphics[scale=0.34]{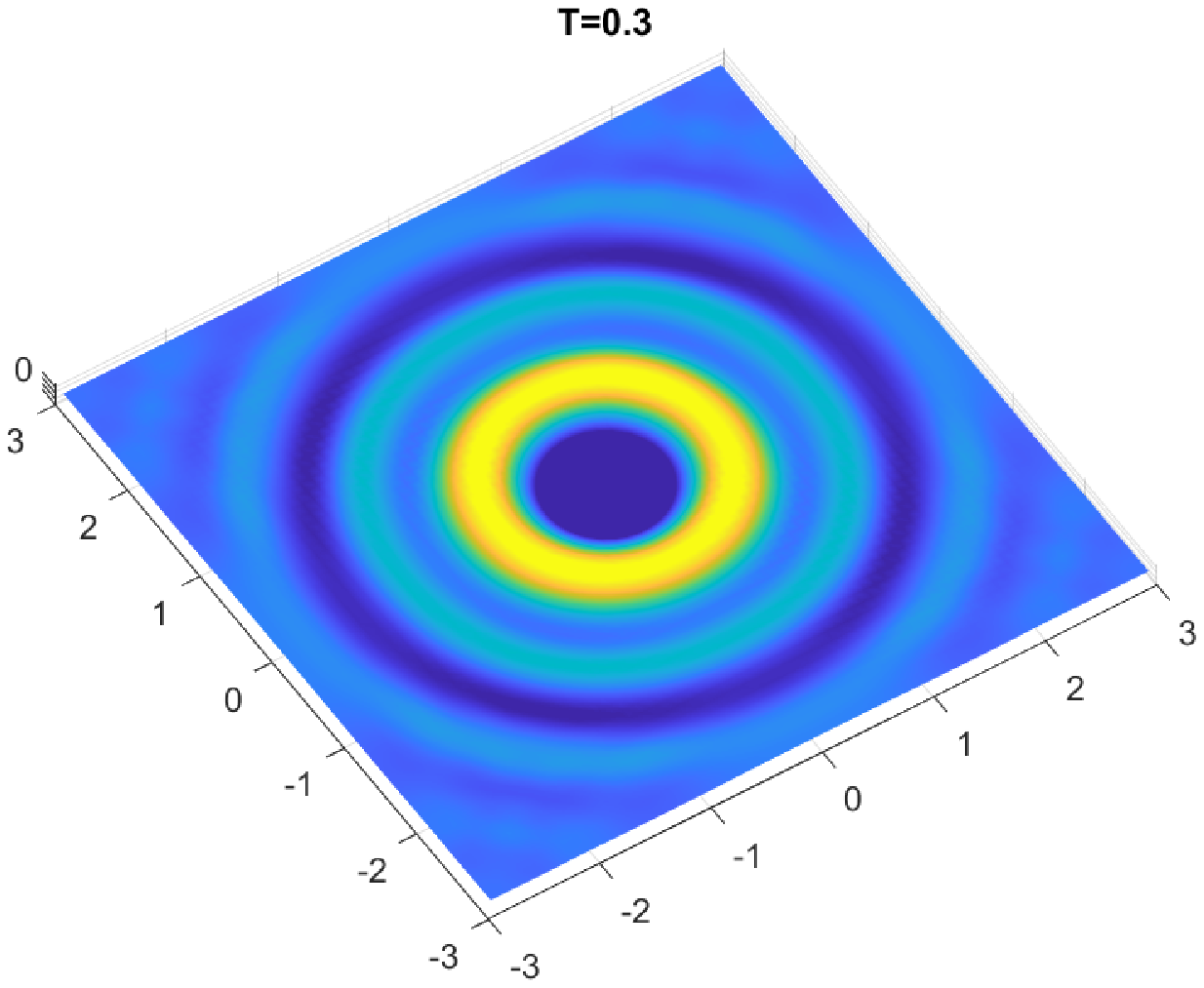}}
  \subfigure[Schr\"odingerization: $T = 0.6$]{\includegraphics[scale=0.34]{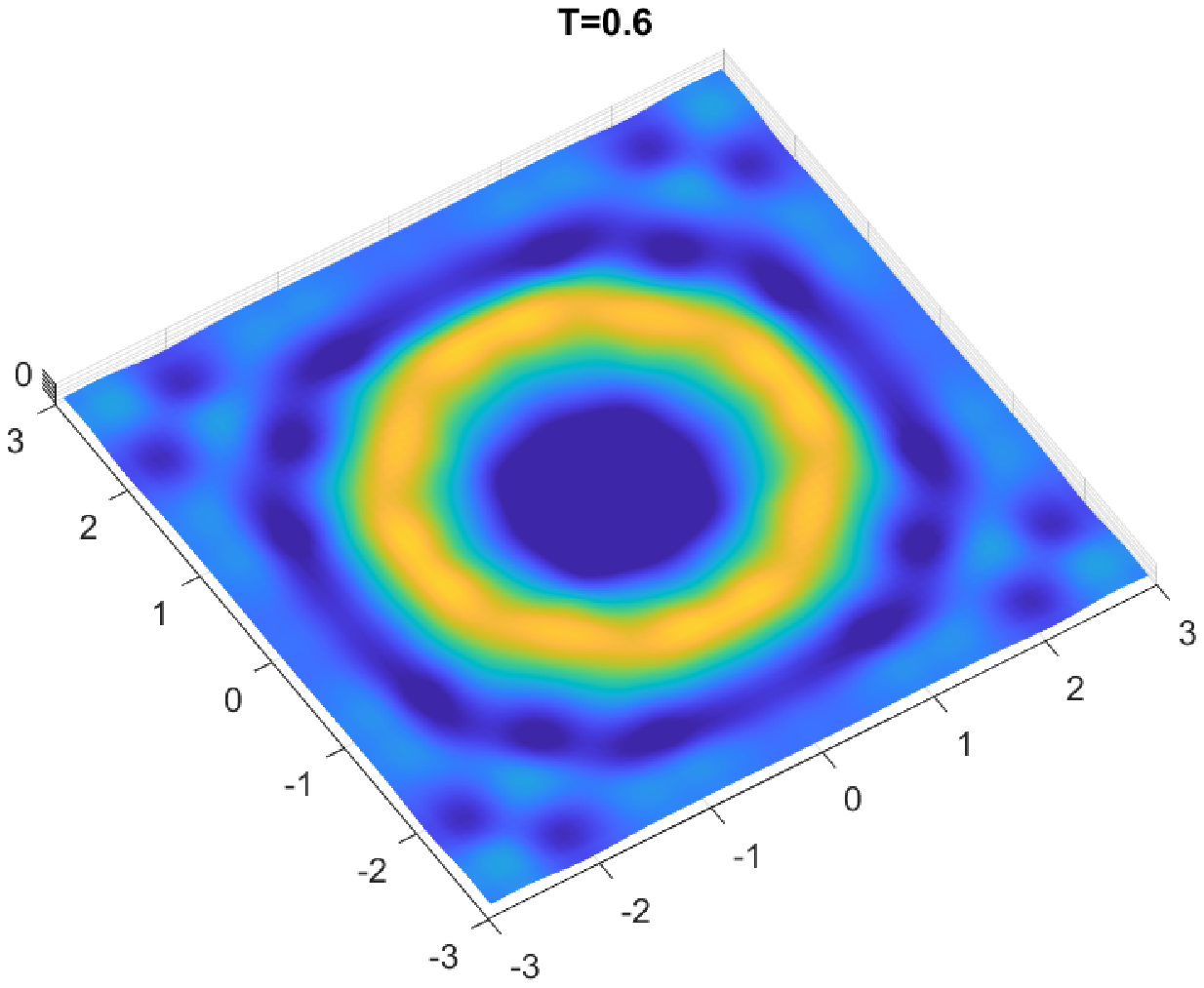}}
  \subfigure[Schr\"odingerization: $T = 0.9$]{\includegraphics[scale=0.34]{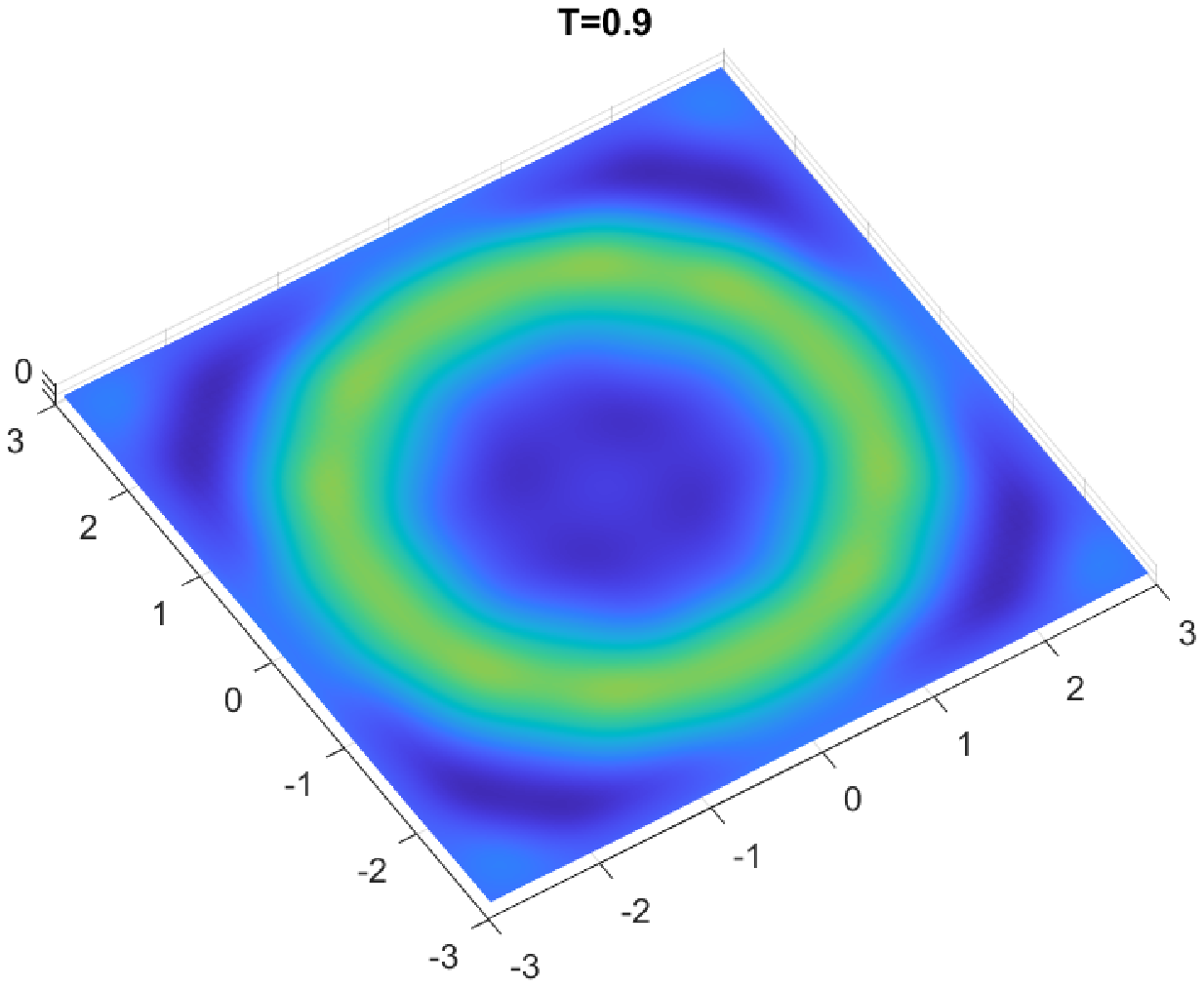}}\\
  \caption{Snapshots of the real parts of $\psi$ computed using the PML method. (a-c): the original form; (d-f): the Schr\"odingerization form.}\label{fig:SchrT0306PML}
\end{figure}

Now we turn to the PML method. We implemented \eqref{PML} on the same test problem, and the results are depicted in Figure \ref{fig:SchrT0306PML}. Here the function $\sigma$ is chosen as
\[\sigma(x) = \begin{cases}
10(x-2.2)^2, & \qquad x>2.2, \\
10(-2.2-x)^2, & \qquad x<-2.2. \\
\end{cases}\]

\medskip

 To implement the DtN approach, one can solve a discrete boundary element equation \cite{wu2020absorbing} to determine the matrix $R$ in \eqref{eq: Rmat}, which then determines the self-energy in \eqref{self-e}.  Since our purpose is to test the Schr\"odingerization approach, we compute $R$ from \eqref{eq: Rmat}  by choosing a relatively large domain $D \supset \Omega$  and simply set $\Omega_\ii = D \backslash \Omega_\i$. For the test example, it suffices to choose $D = [-6,6]^2$.

\begin{figure}[!htb]
  \centering
  \subfigure[Original: $T = 0.3$]{\includegraphics[scale=0.34]{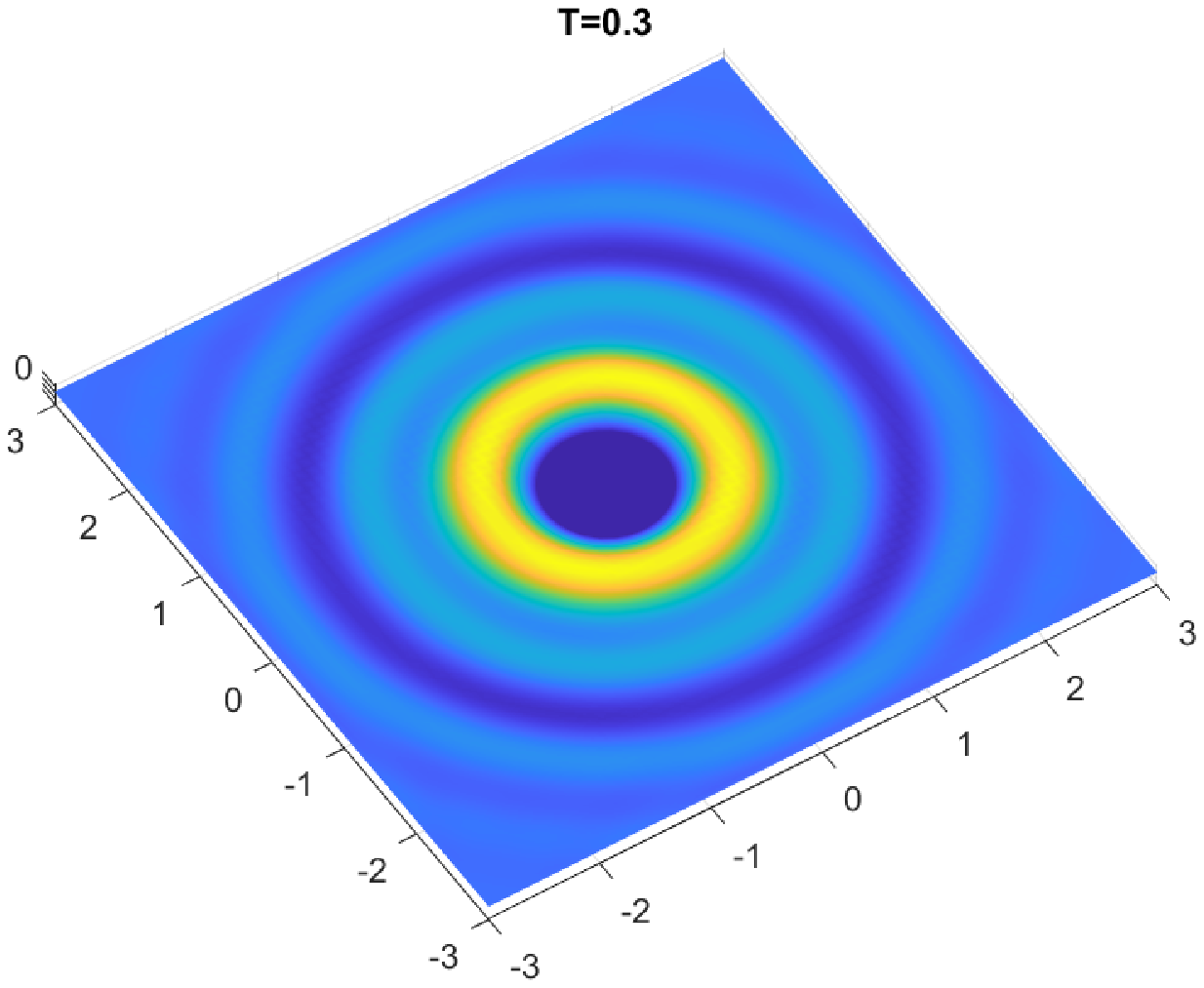}}
  \subfigure[Original: $T = 0.6$]{\includegraphics[scale=0.34]{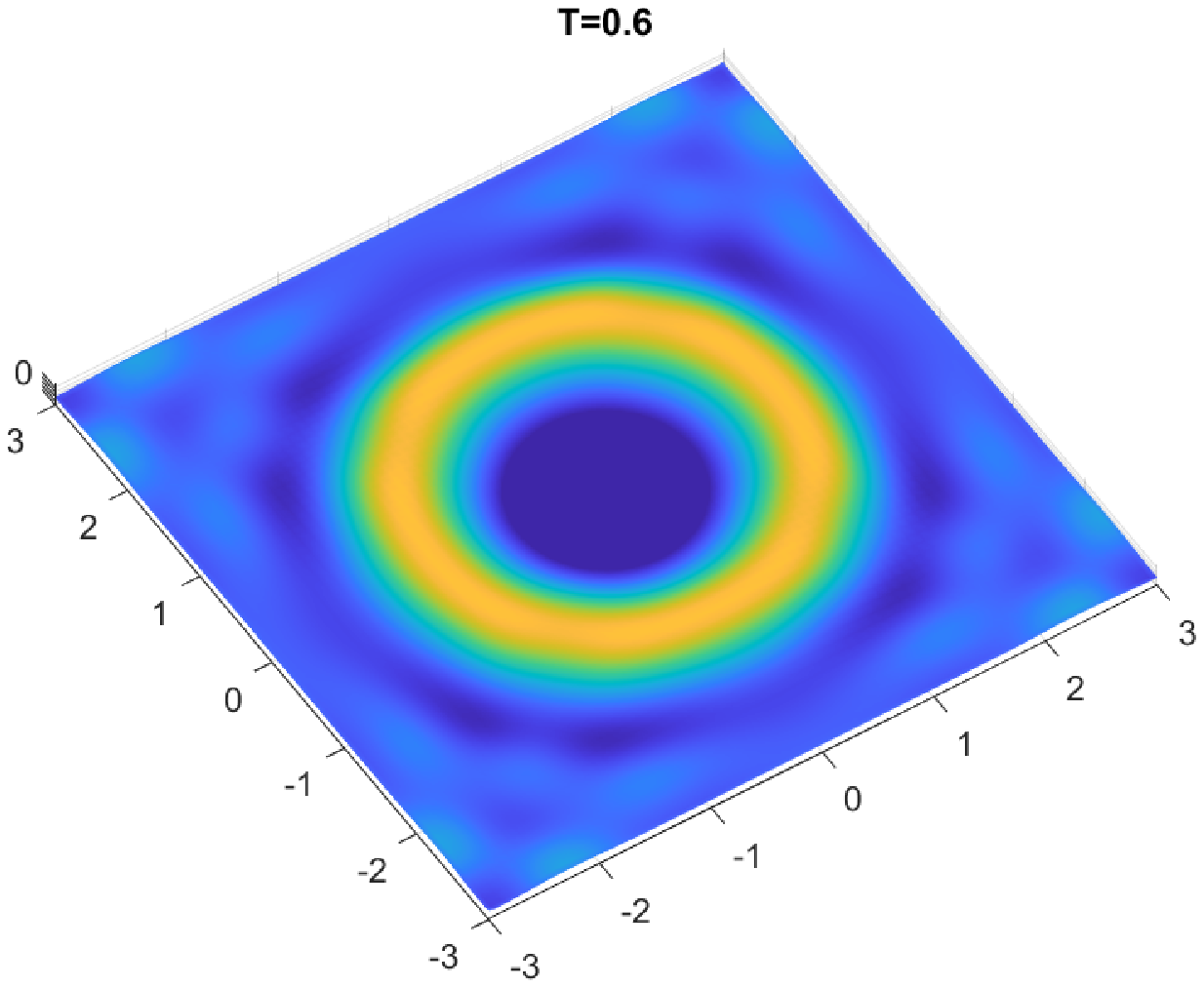}}
  \subfigure[Original: $T = 0.9$]{\includegraphics[scale=0.34]{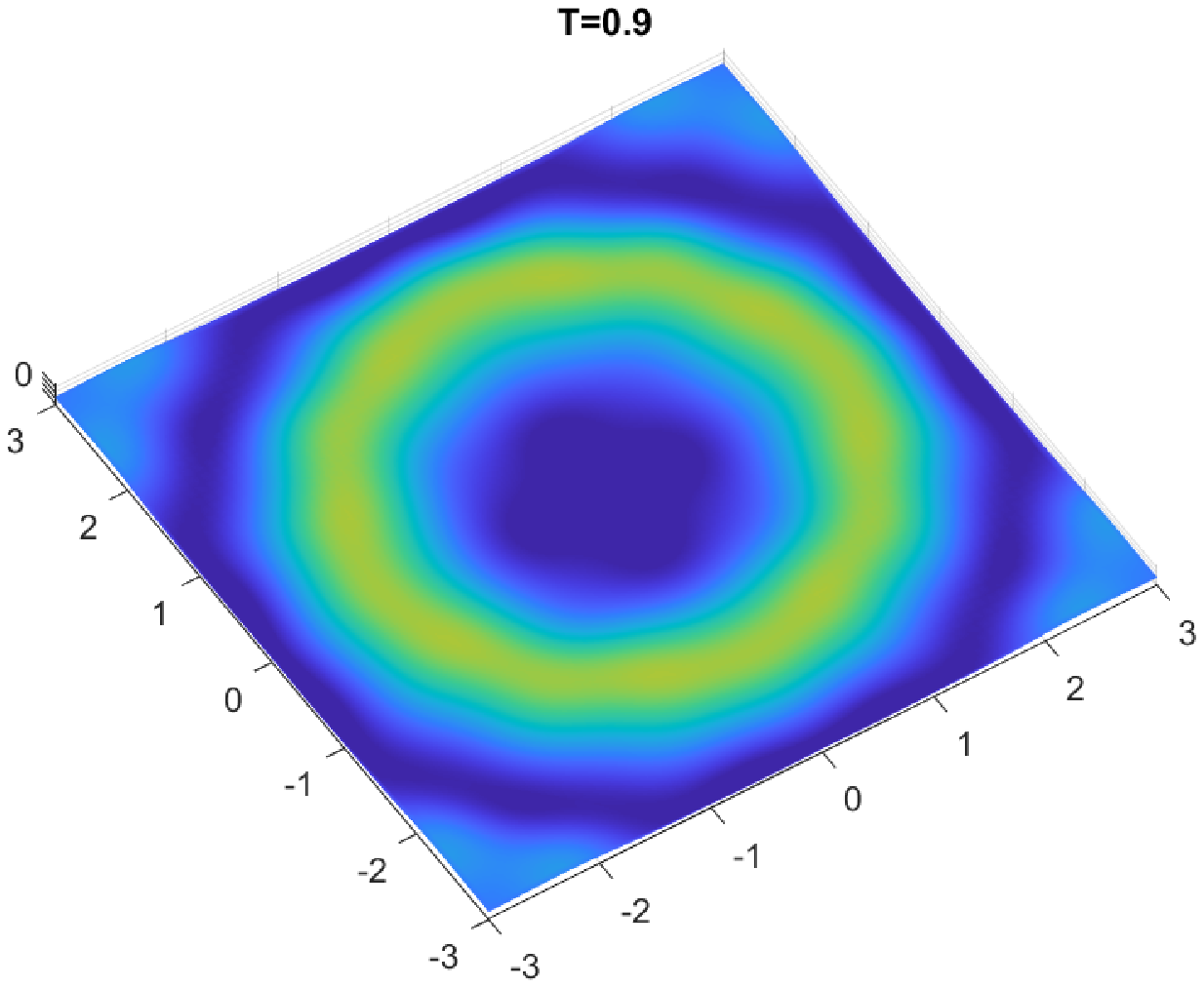}}
  \\
  \subfigure[Schr\"odingerization: $T = 0.3$]{\includegraphics[scale=0.34]{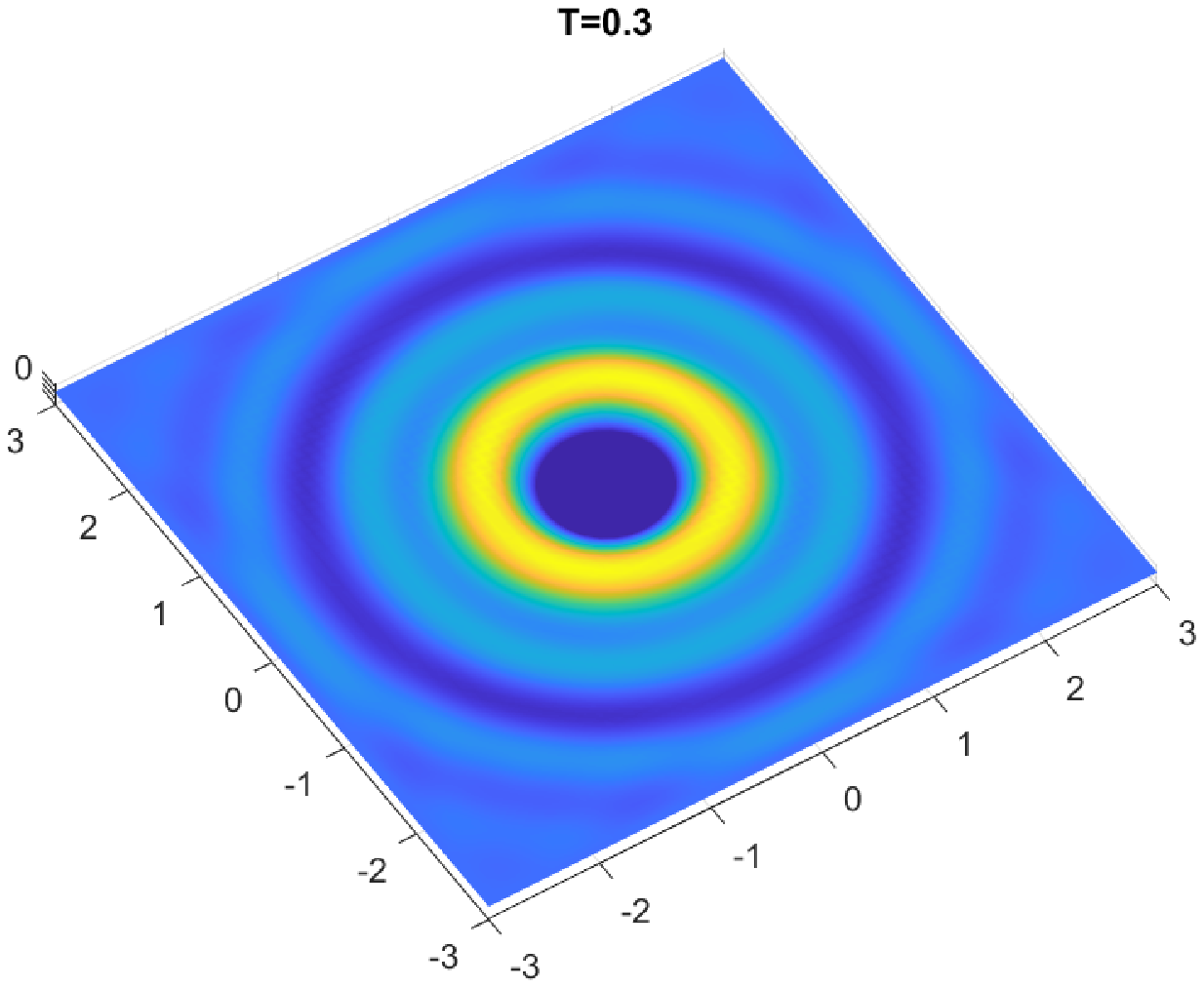}}
  \subfigure[Schr\"odingerization: $T = 0.6$]{\includegraphics[scale=0.34]{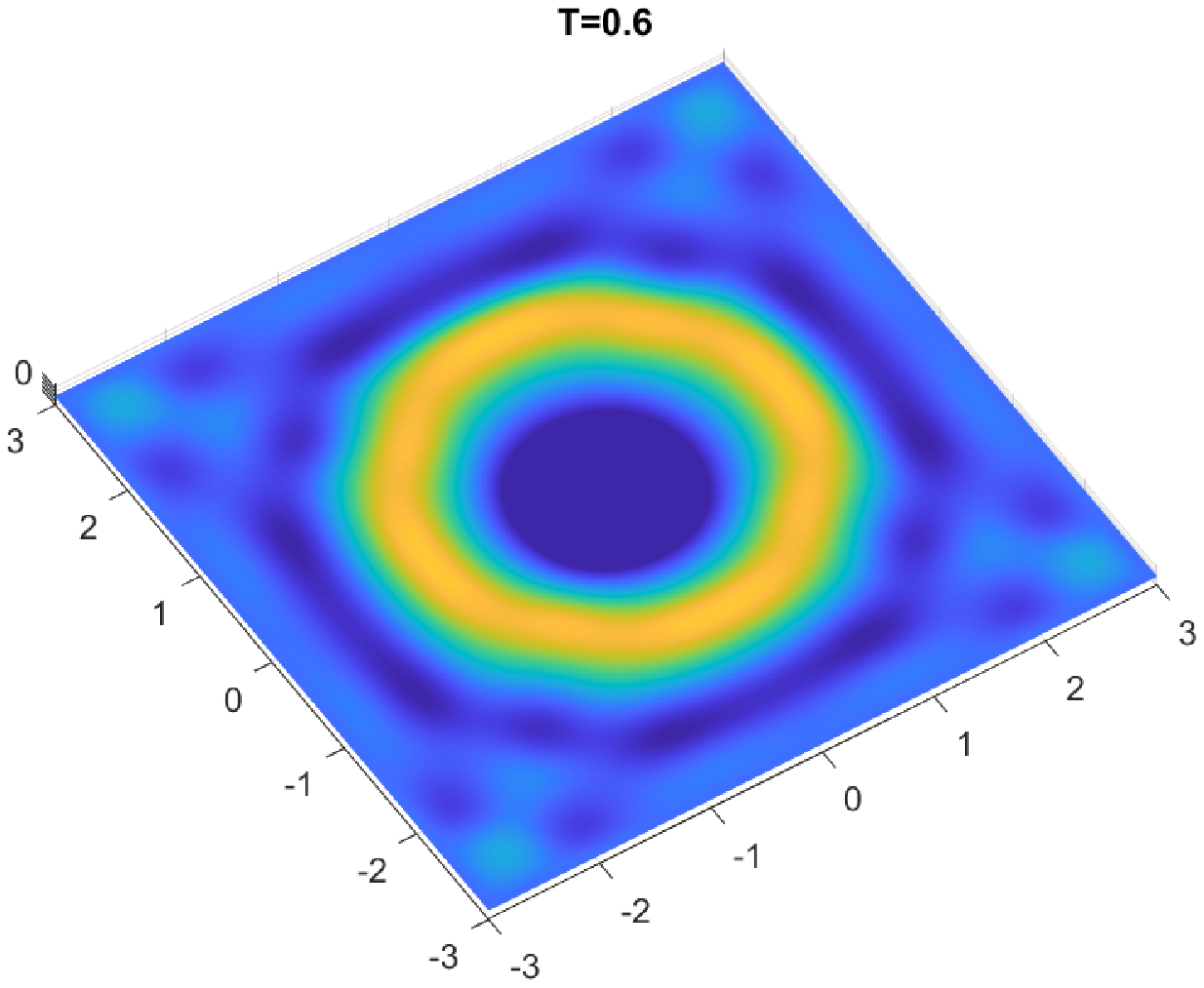}}
  \subfigure[Schr\"odingerization: $T = 0.9$]{\includegraphics[scale=0.34]{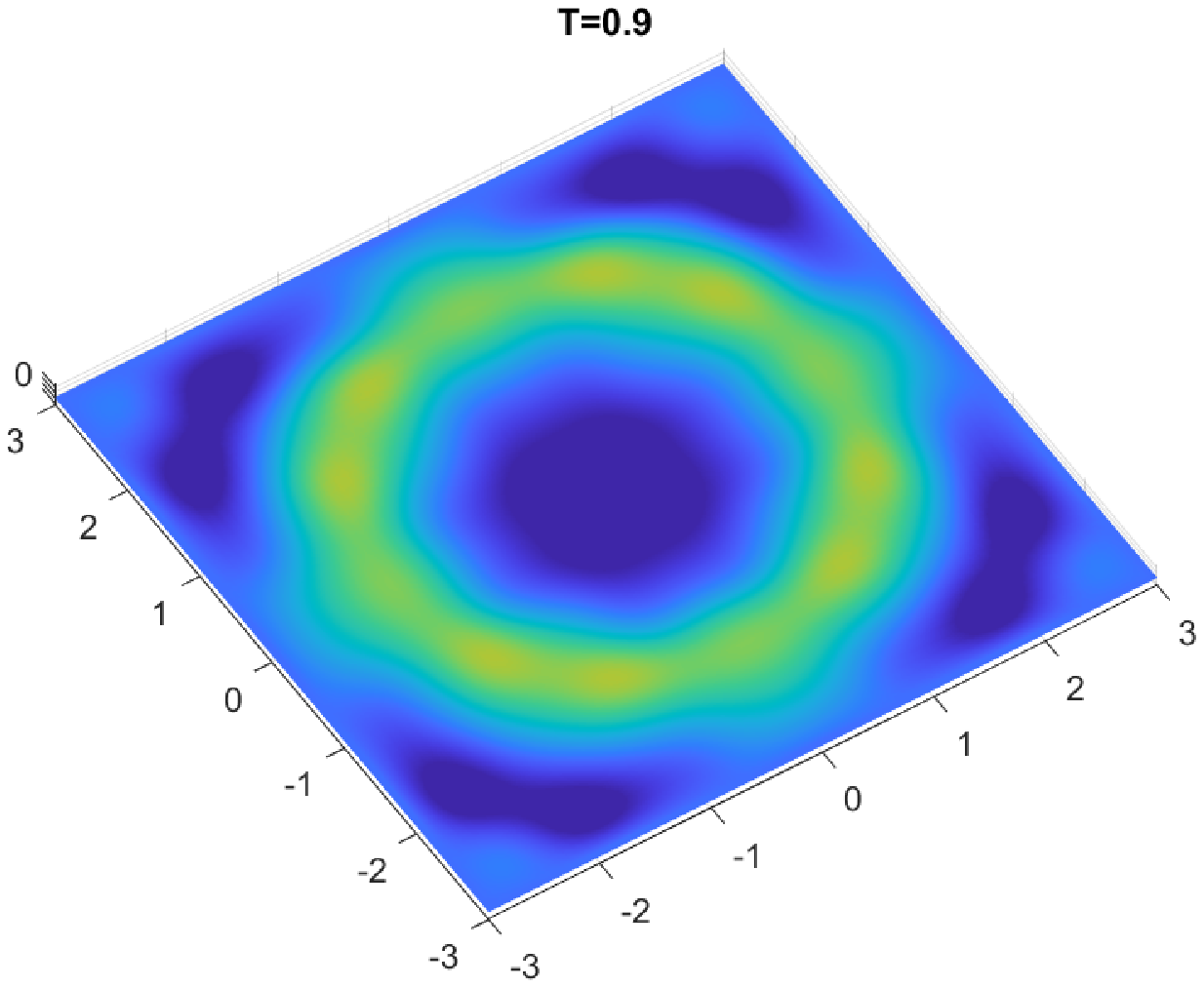}}
  \\
  \caption{Snapshots of the real parts of $\psi$ computed using the DtN method. (a-c): the original form; (d-f): the Schr\"odingerization form.}\label{fig:SchrT0306DtN}
\end{figure}

Since all the matrices ($H_{\i,\i}$, $H_{\i,\ii}$, and $H_{\ii,\ii}$) can be directly extracted from the Hamiltonian matrix on $D$, we can compute the matrices in \eqref{self-e} by direct matrix inversion and multiplications.  In \eqref{DtNm1}, we take $s_0 = 1$. The direct simulation with the DtN ABC can be done by solving the ODEs \eqref{eq:dynamics-AB}.   The results are displayed in the top row  in Figure \ref{fig:SchrT0306DtN}. Meanwhile, a quantum algorithm would solve \eqref{generalSchr} with the two matrices $H_0$ and $H_1 $  from \eqref{H0-H1-dtn}.  We observe that the classical algorithm is able to maintain the  propagating pattern of the wavefront without boundary reflections. The solution based on the Schr\"odingerization  has a similar performance. The slight difference can be attributed by the discretization in the $p$ space.

\section{The implementation complexity of the quantum algorithms}

Once the quantum dynamics with ABCs is turned into a Hamiltonian system \eqref{generalSchr}, one can apply a Hamiltonian simulation algorithm to produce the wave function $\ket{\psi(T)}.$ Due to the possible time-dependence potential $V(\bm x, t)$, a Hamiltonian simulation algorithm for time-dependent Hamiltonian $H(t)$ should be considered. To assess the algorithm complexity associated with the implementation of the ABCs, we use the recent results by Berry et al. \cite[Theorem 10]{berry2020time}, although the other algorithms can also be used for the assessment. Here we simply highlight the query complexity,
\begin{theorem}
    The TDSE \eqref{eq:tdse} with an $s$-sparse Hamiltonian $H(t)$ can be simulated from $t=0$ to $t=T$ within error $\epsilon$ with query complexity,
    \begin{equation}
        \mathcal{O}\left( s \norm{H}_\text{max,1} \frac{\log (s \norm{H}_\text{max,1}/\epsilon )}{ \log\log(\norm{H}_\text{max,1}/\epsilon) } \right).
    \end{equation}
 \end{theorem}

Here the norm is defined as,
\begin{equation}
    \norm{H}_\text{max,1} =\int_0^T \norm{H}_\text{max}(t) dt, \quad
\norm{H}_\text{max}(t):= \max_{i,j} \abs{H_{i,j}(t)}.
\end{equation}

\renewcommand{\arraystretch}{1.8}

\begin{table}[!htb]
\centering
\caption{Summary of the complexity with implementing the time-dependent Schr\"odinger equation \eqref{generalSchr} obtained  from the Schr\"odingerization of the three types of ABCs. The $\widetilde{\mathcal{O}}$ notation rules out logarithmic factors.} \label{tab}
\begin{tabular}{|c|c|}
    \hline
     Artificial Boundary Condition & Query Complexity \\
    \hline
        Complex Absorbing Potential (CAP) & $\widetilde{\mathcal{O}}\left( \frac{T}{\Delta x^2} + \norm{V}_{\text{max},1} + \frac{T \norm{W}_\text{max}}{\Delta x^2} \right)$  \\
    \hline
     Perfectly Matched Layers (PML)  &  $\widetilde{\mathcal{O}}\left( \frac{T}{\Delta x^2} + \norm{V}_{\text{max},1} + \frac{T \norm{\sigma}_\text{max}}{\Delta x^3} \right)$ \\
    \hline
   Dirichlet-to-Neumann Map (DtN) &  $\widetilde{\mathcal{O}}\left(  s_\Sigma \left(
    \frac{ T}{\Delta x^2} + \norm{V}_{\text{max},1}\right) \right)$ \\
     \hline
\end{tabular}
\end{table}

We now discuss the complexity with implementing the time-dependent Schr\"odinger equation \eqref{generalSchr} obtained from the Schr\"odingerization of the three types of ABCs. Since the discretization error for equation \eqref{u2v} is $\mathcal{O}(\Delta p)$ (since $e^{-|p|}$ is only continuous but not continuously differentiable in $p$)  and the discretization in real space has error $\mathcal{O}(\Delta x^2)$, we set
\begin{equation}\label{dpdx2}
    \Delta p = \mathcal{O} \big( \Delta x^2 \big)
\end{equation}
to make the discretization error comparable.\\

\noindent\emph{The CAP method.}  For the CAP method, in light of \eqref{H01CAP}, we obtain directly  that
\[
\begin{aligned}
     \|H_0(t)\|_\text{max}  &\leq \frac{2}{\Delta x^2} + \max_{i,j} |V(x_i,y_j,t)| + \max_{i,j} | \text{Re} W(x_i,y_j)|, \\
    \|H_1\|_{\text{max}} &\leq \max_{i,j} | \text{Im} W(x_i,y_j)|.
\end{aligned}
 \]

Thus we can conclude that
\[\|H\|_\text{max} = \|H_0 \otimes I - H_1 \otimes D_\mu\|_\text{max} \leq  \frac{2}{\Delta x^2} + \norm{V}_\text{max,1} + \frac{\norm{W}_\text{max}}{\Delta x^2},\]
where we have used $\|D_\mu\| = \mathcal{O}(1/\Delta p)$ together with \eqref{dpdx2}.
Meanwhile, the sparsity of $H$ is $s(H) = \mathcal{O}(1)$ (For $d$ dimensions, it is $\mathcal{O}(d)$). Combining these estimates gives the total complexity, as shown in the first row of Table \ref{tab}.

\bigskip

\noindent\emph{The PML method.} We examine the magnitude of the Hamiltonian in \eqref{generalSchr}
from \eqref{PMLLh} and \eqref{SchH},
\begin{align*}
\|H_0(t)\|_\text{max}
& = \|\frac12(L_h-L_h^\dag)\|_\text{max}  \leq \|H_V\|_\text{max}  + \left(\|\bm{\sigma}_x\|_\text{max}  + \|\bm{\sigma}_y\|_\text{max} \right) \left( \|D_x\|_\text{max}  + \|D_y\|_\text{max} \right) \\
& \leq \frac{1}{\Delta x^2} + \max_{i,j} |V(x_i,y_j,t)| + 4\max_{i} |\sigma(x_i)| \frac{1}{\Delta x}.
\end{align*}
Similarly, we have,
\begin{align*}
\|H_1\|_\text{max}
= \|\frac12(L_h+L_h^\dag)\|_\text{max}  \leq (\|\bm{\sigma}_x\| + \|\bm{\sigma}_y\|) ( \|D_x\| + \|D_y\|) \leq 4 \max_{i} |\sigma(x_i)| \frac{1}{\Delta x}.
\end{align*}
These two estimates give
\[\|H\|_\text{max}  \lesssim \frac{1}{\Delta x^2} +  \norm{V}_\text{max} + \norm{\sigma}_\text{max}  \frac{1}{\Delta x \Delta p}. \]
Finally, one can also deduce  from \eqref{PMLLh} and \eqref{SchH} that the sparsity $s(H) = \mathcal{O}(1)$. The total complexity is shown in the second row of Table \ref{tab}.

\bigskip

\noindent\emph{The DtN method.} The query complexity with the DtN method is provided in the third row of Table \ref{tab}. To prove this estimate, we  follow the equations \eqref{self-e} and \eqref{H0-H1-dtn}. First, we notice that $S=H_{\iiicii} H_{\iiciii}$ in \eqref{self-e} is sparse and the elements are of the order $\frac{1}{\Delta x^4}.$ Therefore, $ \norm{\Sigma_{\i,\Gamma}}_\text{max}= \mathcal{O}\left(\frac{1}{\Delta x^2}\right).$
Following the same observation, we deduce that  $\norm{R}_\text{max} =\mathcal{O}\left(\frac{1}{\Delta x^2}\right).$ Using the formulas \eqref{G0G1} that we derived in the previous section, we find that,
$
\norm{G_0}_\text{max} = \mathcal{O}\left(\frac{1}{\Delta x^2}\right),$ but $\norm{G_0}_\text{max} = \mathcal{O} (1).$

Inserting these estimates into \eqref{H0-H1-dtn}, we have
\[\|H(t)\|_\text{max} = \mathcal{O} \left( \frac{1}{\Delta x^2} + \norm{V}_\text{max}
+ \frac{1}{\Delta p} \right).\]

\begin{figure}[!htb]
  \centering
  \includegraphics[scale=0.48]{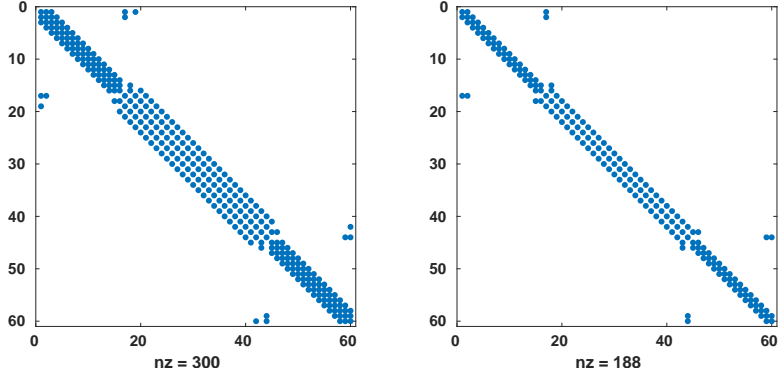}\\
  \caption{The effective sparsity of $\Sigma_{\Gamma,\Gamma}$. Left: entries satisfying $\abs{\Sigma_{ij}}> 0.01 \norm{\Sigma_{\Gamma,\Gamma} }_{\text{max}}$ (one-percent cut);  Right: entries satisfying $\abs{b_{ij}}> 0.02 \norm{\Sigma_{\Gamma,\Gamma} }_{\text{max}}$ (two-percent cut). $\norm{\Sigma_{\Gamma,\Gamma} }_{\text{max}}:= \max_{i,j}\abs{\Sigma_{ij}}.$ }\label{fig:sparsityB}
\end{figure}

The other factor that contributes to the complexity is the sparsity of the matrices in \eqref{self-e}. Numerical observations of these matrices suggest that they are close to identity matrices. For example, for the matrix $\Sigma_{\Gamma,\Gamma}$ in \eqref{self-e}, we plotted in Figure \ref{fig:sparsityB} its entries by removing entries that are much smaller than its maximum norm.
We let $s_\Sigma$ be such an effective sparsity and the overall complexity is linear in $s_\Sigma$.

 \section{Conclusion}
 In this article, we use the Schr\"odingerization approch recently introduced in \cite{jin2022quantumshort, jin2022quantum} for quantum dynamics with articifial boundary conditions (ABCs).
 While a quantum dynamics with ABC is no longer a Hermitian Hamiltonian system, which is most natural for quantum simulation, the Schr\"odingerization approach makes it so in a simple fashion.
 We give the implementation details and estimate the computational complexities for three representative ABCs, including  the complex absorbing potential method, perfectly matched layer methods, and Dirichlet-to-Neumann approach.
 Our numerical experiments validate this apporach, demonstrating that the Schr\"odingerized systems yield the same results as the original dynamics with ABCs.

 Our approach provides a simple quantum simulation algorithms for quantum dynamics in unbounded domains. As pointed out in \cite{jin2022quantumshort, jin2022quantum}, the Schr\"odingerization apporach is not only applicable to quantum dynamics with ABCs, it also applies to general linear partial differential equations with physical boundary conditions. It can also be applied to such equations with interface conditions. These will be the subject of our future research.

 \section*{Acknowledgements}
SJ was partially supported by the NSFC grant No.~12031013, the Shanghai Municipal Science and Technology Major Project (2021SHZDZX0102), and the Innovation Program of Shanghai Municipal Education Commission (No. 2021-01-07-00-02-E00087).  NL acknowledges funding from the Science and Technology Program of Shanghai, China (21JC1402900). YY was partially supported by China Postdoctoral Science Foundation (no. 2022M712080). XL has supported by a Seed Grant from the Institute of Computational and Data Science (ICDS) at Penn State.

\bibliographystyle{alpha}

\newcommand{\etalchar}[1]{$^{#1}$}

\end{document}